\documentclass[letterpaper]{article}
\usepackage{amsfonts}
\usepackage{amsthm}
\usepackage{amsbsy}
\usepackage{amssymb}
\usepackage{epsfig}
\usepackage{rotating}

\usepackage[all]{xy}
\usepackage{amsfonts}
\usepackage{amsmath}
\usepackage{eucal}
\usepackage{amssymb}

\newtheorem{theorem}{Theorem}[section]

\newcommand{\eq}{\begin{equation}}
\newcommand{\feq}{\end{equation}}
\newcommand{\eqn}{\begin{eqnarray}}
\newcommand{\feqn}{\end{eqnarray}}
\newcommand{\arr}{\begin{eqnarray*}}
\newcommand{\farr}{\end{eqnarray*}}

\newcommand{\cHH}{\mathfrak{h}}
\newcommand{\cP}{\mathfrak{p}}
\newcommand{\Tr}{{\rm Tr}}

\begin{document}
\title{Exceptional groups, symmetric spaces and applications}

\author{Sergio L.~Cacciatori\footnote{Dipartimento di Fisica e Matematica,
Universit\`a dell'Insubria Milano, 22100 Como, Italy, and
I.N.F.N., sezione di Milano, Italy.
E-mail address: sergio.cacciatori@uninsubria.it} and
B.~L.~Cerchiai\footnote{Lawrence Berkeley National Laboratory, Theory Group,
Bldg 50A5104, 1 Cyclotron Rd, Berkeley CA 94720-8162, USA and
University of California Berkeley, Center for Theoretical Physics,
366 LeConte Hall \#7300, Berkeley, CA 94720-7300, USA.
E-mail address: BLCerchiai@lbl.gov}}

\maketitle

\vspace{-15em}
\hfill{UCB-PTH-09/17}

\hfill{LBNL-1938E}\\[15em]

\begin{abstract}
In this article we provide a detailed description of a technique
to obtain a simple parametrization for different exceptional
Lie groups, such as $G_2$, $F_4$ and $E_6$, based on their fibration
structure. For the compact case, we construct
a realization which is a generalization of the Euler angles
for $SU(2)$, while for the non compact version of
$G_{2(2)}/SO(4)$ we compute the Iwasawa decomposition.
This allows us to obtain not only an explicit expression
for the Haar measure on the group manifold, but
also for the cosets $G_2/SO(4)$, $G_2/SU(3)$, $F_4/Spin(9)$, $E_6/F_4$ and
$G_{2(2)}/SO(4)$ that we used to find the concrete realization
of the general element of the group. Moreover, as
a by-product, in the simplest case of $G_2/SO(4)$, we have
been able to compute an Einstein metric and the vielbein.

The relevance of these results in physics is discussed.
\end{abstract}

\section{INTRODUCTION}
In this article we describe our technique to analyze the
structure of exceptional Lie groups, which is based on
constructing a generalized Euler parametrization by starting
from a suitable fibration. We review our results on $G_2$
\cite{CCdVOS, C}, $F_4$ \cite{BCCS1} and $E_6$ \cite{BCCS2}.
We also provide some new insights on the geometry of the
non compact versions of these groups, by using the Iwasawa
decomposition, and in particular we apply it to $G_{2(2)}$.
Our method allows us to explicitly calculate the Haar measure
for the group manifold, and, as it is compatible with the fibration
used to compute it, it naturally provides a metric for the
corresponding coset as well.

The layout of this paper is as follows. In section 2 we recall
some of the basic facts about Lie groups and Lie algebras, that we need
later. In section 3 we explain in detail how the
generalized Euler parametrization is defined and we study some toy
model to exemplify it. Then in the following sections we apply it to
different exceptional Lie groups.
In section 4 we construct $G_2$ in two different ways as a fibration,
first with $SU(3)$ as a fiber and then with $SO(4)$ as a fiber.
In section 5 we determine the $Spin(9)$ Euler angles for $F_4$,
which we then use in section 6 to obtain the $F_4$ Euler angles
for $E_6$. Finally, in section 7 we introduce the Iwasawa decomposition
for the non compact version of the Lie groups, which we then apply
to $G_{2(2)}$ in section 8.

Since we are able to get an explicit expression for the Haar measure
on the group manifold, the most immediate application of our results
is the possibility of evaluating integrals \cite{KS}.
Until now the only available method to compute some of them was to use
the invariance properties of the Haar
measure, but knowing its explicit form gives an analytic way to
calculate many of them directly.

In physics exceptional Lie groups appear naturally as the symmetry (gauge)
groups of field theories which are low energy limits of certain heterotic
string models \cite{FW}. Besides from being relevant for string
phenomenology, these theories are interesting by themselves, e.g.
$E_6$ as a candidate for the symmetry group in a grand unified theory
of high energy physics \cite{OR} and $G_2$ as a possible
example of a non confining gauge theory \cite{P}.
While the local properties of a field theory are determined exclusively at
the level of the corresponding Lie algebra, in order to obtain
non-perturbative results it is necessary to make use of the full global
structure of the Lie group, because of the need for evaluating integrals
on the group manifold.
Being able to solve them analytically has drastically reduced the
computer power required to run a lattice simulation. For instance
our expressions for $G_2$ are the base for the Montecarlo analysis
presented in \cite{CEGLP}.

Moreover, our technique can also be applied to the noncompact versions
of the Lie groups, such as $G_{2(2)}$, $F_{4(4)}$, $E_{6(6)}$ or
$E_{7(7)}$. In this case, another parametrization is the
Iwasawa decomposition. As its construction uses a nilpotent subalgebra,
it is particularly simple and is therefore very useful. In physics these
groups represent the U-duality of supergravity theories in different
dimensions.

One of the most interesting features of our method is that it is based
on identifying a suitable subgroup and in studying the corresponding
fibration. As a consequence it automatically yields an explicit expression
for the coset space as well as for its metric, measure and vielbein, since
the geometry on the group induces a geometry on the base. In the case of
the maximal compact subgroups of noncompact exceptional Lie groups,
e.g. $SO(4)$ for $G_{2(2)}$ or $SU(8)$ for $E_{7(7)}$, these symmetric
spaces turn out to be Einstein spaces. Being solutions of Einstein
equations, they are relevant by themselves for general relativity.

In supergravity some of these cosets are interpreted as the scalar
fields of the associated sigma model \cite{FM}. Moreover, they
can represent the charge orbits of black holes when the attractor mechanism
is studied \cite{GNPW} and they also appear as the moduli spaces for black
holes. In \cite{CFMZ} they are used to investigate the deep connection
between black holes properties, duality and supergravity.

As an example, the coset space $G_{2(2)}/SO(4)$ studied in section 8
is relevant for black ring solutions in 5-dimensional supergravity
\cite{CW}.

Finally, these symmetric spaces can be used to describe the entanglement
of qubits and qutrits in information theory \cite{S}.

\section{GENERAL SETTINGS}
Because of their importance for the rest of the chapter and in order to set
our conventions, we recall here some basic facts about semisimple Lie
groups (see \cite{FH}).

\subsection{Lie algebras from Lie groups}

A Lie group $G$ is a group which is also a differential manifold and for
which the group structure and the differential structure are compatible.
This means that the two basic group operations, the
product and the inversion, are required to be differentiable maps with
respect to the differential structure. The dimension of the group
is the dimension of $G$ as a manifold. Here we consider only finite
dimensional groups. In this case the differentiability of the inverse map
is a consequence of the differentiability of the product map and of
the implicit function theorem. We use the symbol $e$ for the unit element,
which therefore identifies a particular point on $G$.\\
For any $g\in G$ we can define two maps:
\begin{eqnarray*}
&& L_g: G\longrightarrow G, \ h\mapsto gh,\\
&& R_g: G\longrightarrow G, \ h\mapsto hg,
\end{eqnarray*}
called the left and the right translation respectively. Note that with
respect to the composition product, $L_g$ and $R_{g'}$ commute.
They define a left and a right action of the group on itself:
\begin{eqnarray*}
&& L: G\times G \longrightarrow G;\ (g,h)\mapsto L_g(h),\\
&& R: G\times G \longrightarrow G;\ (g,h)\mapsto R_g(h).
\end{eqnarray*}
Note that $L_g$ and $R_g$ are not homomorphisms. A homomorphism associated
to these actions is:
\begin{equation}
\phi_g: G\longrightarrow G;\ h\mapsto R_{g^{-1}} L_g h =ghg^{-1}.
\end{equation}
Differentiating the $L_g$ map at the identity, we have
$$
(dL_g)_e : T_e G \longrightarrow T_g G.
$$
This operation associates to each vector $\xi\in T_eG$ a non vanishing
vector field $X_\xi$
$$
X_\xi : G\longrightarrow TG;\ g\mapsto (dL_g)_e (\xi) \in T_g G,
$$
which is well defined globally.
Note that $X_\xi (e)=\xi$.
In this way, given a basis $\{\tau_1,\ldots,\tau_n\}$ of $T_e G$, at each
point $g$ we can obtain a set of vector fields which determine a basis
for $T_gG$. This shows that the tangent bundle of $G$ is trivial.\\
An important property of the field $X_\xi$ is that it is $L_g$-invariant
({\it left invariant}). This means $(L_g)_*  X_\xi =X_\xi$. Viceversa,
given a left invariant vector field $V$, it can be verified that $V(e)\in T_e G$
and $V=X_{V(e)}$.
Thus, the left invariant vector fields form a finite dimensional vector
space ${\cal X}^L(G)\simeq T_eG$. Moreover, ${\cal X}^L(G)$ is closed
under the Lie bracket of vector fields:
$$
[X,Y]\in {\cal X}^L(G), \quad\ \mbox{ for all } \quad\ X,Y\in {\cal X}^L(G) \quad\ ([X,Y]={\cal L}_X Y),
$$
where ${\cal L}_X$ is the Lie derivative along $X$. Thus,
$$
\mathfrak{g}\equiv Lie(G):= \{{\cal X}^L(G) , [,] \}
$$
defines an algebra: the {\it Lie algebra associated to G}. The Lie
product has the properties of being antisymmetric and of satisfying
the Jacobi identity.

\subsection{Adjoint representations and the Killing form}

A powerful way to ``describe'' the structure of a group is by means of
its representations.
A representation of a group $G$ on a vector space $V$ (real or complex) is
a homomorphism $r:G\longrightarrow {\rm Aut} (V)$, where ${\rm Aut} (V)$
is the group of automorphisms of $V$ with the composition as product.
A representation is irreducible if $V$ does not admit any proper invariant
subspaces, and it is faithful if
${\rm Ker} (r)=e$.
In a similar way, a representation of a Lie algebra on $V$ is a
homomorphism $\rho: \mathfrak{g}\longrightarrow {\rm End}(V)$,
where ${\rm End} (V)$ is the Lie algebra of endomorphisms of $V$ with
the bracket of operators as Lie product. Noting that
$End(V)=Lie(Aut(V))$ and identifying $Lie(G)$ with $T_eG$, it can be seen
that a representation of the algebra can be obtained from a representation
of the group by differentiation: $\rho=dr_e$.\\
Among the representations of a group, an example which can be constructed
in a natural way is the Adjoint. It is the representation over the Lie
algebra $V=\mathfrak{g}$ obtained in the following way through the homomorphism
$\phi_g$ introduced above.
For any fixed $g$, we define the map:
\begin{equation}
Ad_g : T_eG \longrightarrow T_eG; \quad Ad_g:= (d\phi_g) \nonumber
\end{equation}
where $d$ is the differential of $\phi_g$ at the identity.
Then the {\it Adjoint representation} of the group is defined by:
\eq
Ad: G\longrightarrow {\rm Aut}(T_eG);\quad g\mapsto Ad_g.
\feq
Differentiating at the identity yields the {\it adjoint representation}
of the Lie algebra
\eq
ad: \mathfrak{g}\longrightarrow {\rm End}(T_eG);\quad a\mapsto ad_a,
\feq
where $ad_a (b)=[a,b]$ for all $b\in \mathfrak{g}$.\\
Next, from the adjoint representation of the algebra, the Killing form on
$\mathfrak{g}$ can be constructed as follows:
\eq
K: \mathfrak{g}\times \mathfrak{g}\longrightarrow \mathbb{K};\ (a,b)
\mapsto K(a,b):=\Tr (ad_a ad_b),
\feq
where $\mathbb{K}=\mathbb{R}, \mathbb{C}$ is the field of $\mathfrak{g}$.
The Killing product is symmetric and $ad$-invariant, which means:
$$
K(ad_a(b),c)+K(b, ad_a(c))=0.
$$
This defines a symmetric two form over $T_eG$ which in turn, using the left
translation, induces a symmetric two form over the whole group:
\eq
K_G : G\longrightarrow T^*G\otimes T^*G;\ g\mapsto L_{g^{-1}}^* K,
\feq
the pullback of $K$ under $L_{g^{-1}}$. In general the Killing form is degenerate.
It is obviously left invariant.
If we choose a basis $\{\tau_i\}$ for $\mathfrak{g}$ and define the
corresponding structure constants $f_{ij}^{\ \ k}$ as
$[\tau_i, \tau_j]=\sum_{k=1}^n f_{ij}^{\ \ k}\tau_k$, then the Killing form
turns out to be $K_{ij}=K(\tau_i,\tau_j)=\sum_{l,m} f_{il}^{\ \ m}f_{jm}^{\ \ l}$.
The $ad$-invariance implies that the covariant tensor
$f_{ijk}:=f_{ij}^{\ \ l}K_{lk}$ is totally antisymmetric.
In the basis $\{\mu^i\}$ which is canonically dual
to $\tau_j$, (i.e. $\mu^i(\tau_j)=\delta^i_j$), the Killing form takes the
particularly simple expression $K=\sum_{ij} K_{ij} \mu^i \otimes \mu^j$.\\
Finally, an important role is played by the Cartan 1-form. It is a Lie
algebra valued form defined as
$J:= \sum_{i}(L_g)_* (\tau_i \mu^i)=\sum_{i}J^i X_{\tau_i}$
and it can be used to rewrite the Killing form as
$K_G=K(J,J)=\sum_{ij} J^i\otimes J^j K_{ij}$.

\subsection{Simple Lie algebras classification}

Starting from a finite dimensional Lie group, the associated Lie algebra
can be easily determined. Being a linear space, it is much easier
to analyze than the group itself.
There is a very interesting class of Lie algebras, which are completely
classified: the semisimple Lie algebras.
A {\it semisimple Lie algebra} is a Lie algebra of dimension higher than $1$, which does
not admit any Abelian proper ideals. If it does not contain any proper ideal
at all, it is called a {\it simple Lie algebra}. It can be shown that any 
semisimple Lie algebra can be written as a direct sum of simple algebras
in a unique manner (up to isomorphisms). An important result is that a
Lie algebra is semisimple if and only if the corresponding Killing form
is non degenerate.\\
{From} the definition, it follows that for a semisimple algebra ${\rm Ker}(ad)=0$,
so that the adjoint representation is faithful. The Lie algebra can
then be identified with its adjoint representation. This allows a classification
of all complex (finite dimensional) simple Lie algebras by performing a
classification of their adjoint representation. As the main ingredients will
be used later, let us recall the main steps.\\
Any simple Lie algebra contains a unique (up to isomorphisms) Cartan
subalgebra, a maximal Abelian $\mathfrak{h}\subset \mathfrak{g}$
subalgebra such that for each $h\in \mathfrak{h}$, $ad_h$ is diagonalizable.
Then $r={\rm dim} (\mathfrak{h})$ is called the rank of $\mathfrak{g}$.
All such operators $ad_h$ are simultaneously diagonalizable and their
eigenvalues are called the roots $\alpha\in\mathfrak{h}^*$ of the algebra:
$$
ad_h (\lambda_\alpha)=\alpha(h) \lambda_\alpha, \quad\ 0\neq \lambda_\alpha \in \mathfrak{g}.
$$
Since $\mathfrak{g}$ is finite dimensional, the set of all roots $Root(\mathfrak{g})$ is finite. If $\Lambda_\alpha$ is the eigenspace of $\alpha$
then $0$ is a root, $\Lambda_0=\mathfrak{h}$, and $\mathfrak{g}=\bigotimes_{\alpha\in Root(\mathfrak{g})} \Lambda_\alpha$, with the properties that
$[\lambda_\alpha,\lambda_\beta]\in \Lambda_{\alpha+\beta}$, and that it
vanishes if $\alpha+\beta$ is not a root. \\
{From} the $ad$-invariance it follows that
$K(\lambda_\alpha,\lambda_\beta)=0$
if $\alpha+\beta\neq 0$. It can also be shown that if $\alpha$ is a
non vanishing root, then $k\alpha$ is a root if and only if 
$k=0,\pm1$ and ${\rm dim}(\Lambda_\alpha)=1$.
If $K_C$ is the restriction of the Killing form to the Cartan subalgebra,
it follows that $K_C$ is non degenerate and therefore it defines a natural
isomorphism between $\mathfrak{h}$ and $\mathfrak{h}^*$, as well as a 
bilinear form $(|)$ on $\mathfrak{h}^*$ in an obvious way.
It also follows that $Root(\mathfrak{g})$ is real, in the sense that it
contains a basis for $\mathfrak{h}^*$, such that the remaining roots are
real combinations of the basis elements and that it is possible to 
consistently define the $r$-dimensional real space 
$\mathfrak{h}^*_{\mathbb{R}}=\langle Root(\mathfrak{g}) \rangle_{\mathbb{R}}$.
Up to a multiplicative constant, $(|)$ defines a Euclidean scalar product 
on $\mathfrak{h}^*_{\mathbb{R}}$. The main result we need in this 
context is:\\
{\bf The Cartan Theorem:} If $\alpha$ and $\beta$ are two non vanishing roots, then $n_{\alpha\beta}:=2\frac {(\alpha|\beta)}{(\alpha|\alpha)}\in \mathbb{Z}$
and $\beta-n_{\alpha\beta} \alpha$ is also a root (Weyl reflection). \\
This strongly constrains the relations among the roots, because if
$|\alpha|$ and $\theta_{\alpha\beta}$ are respectively the norm and 
the angle between two roots defined by the Euclidean scalar product, then
$$
\frac {|\beta|^2}{|\alpha|^2}=\frac {n_{\alpha\beta}}{n_{\beta\alpha}}, \qquad\ \cos^2 \theta_{\alpha\beta}=\frac 14 n_{\alpha\beta}n_{\beta\alpha}.
$$
At this point it is clear that all the information on the algebra is 
contained in the root system. A simple root system $SR$ is defined as a
basis of the root space, such that all the remaining roots are combinations
of $SR$ with integer coefficients of the same sign.
Such a system always exists, even though in general it is not unique, 
and it decomposes the root space into a positive and a negative part:
$Root(\mathfrak{g})=R^+\oplus R^-$.
Given a simple root system $SR=\{\alpha_1,\ldots,\alpha_r\}$,
the numbers $n_{ij}$ associated to it by the Cartan Theorem must
be all non positive if $i\neq j$, while for $i=j$ it has to be $n_{ii}=2$.
Moreover, either $|n_{ij}|$ or $|n_{ji}|$ is always 1 if $i\neq j$.
These numbers characterize $SR$ completely (up to obvious equivalences)
and define the the Cartan matrix $C_{ij}=n_{ij}$, which has the 
properties: 
$C_{ii}=2$,
$C_{ij}\leq 0$ and
$C_{ij}\neq0$ if and only if $C_{ij}\neq0$, $i\neq j$. 
To classify all the simple Lie algebras, it is therefore enough to
classify all the $SR$ systems, or, equivalently, all the Cartan matrices 
compatible with them. This is done graphically by means of the Dynkin
diagrams: A dot $\circ$ is associated to each of the $r$ simple roots.
Two roots are then connected by $N_{ij}=n_{ij} n_{ji}$ lines with a 
$>$ indicating the direction from the longer root to the shorter one. 
Simple algebras correspond to a connected Dynkin diagram. It turns out
that the admissible Dynkin diagrams can be classified into four classical 
series: $A_r, B_r, C_r, D_r$, $r$ being the rank
of the corresponding algebras, plus five exceptional cases: 
$G_2$, $F_4$, $E_6$, $E_7$, $E_8$.
The corresponding Dynkin diagrams can be found for example in \cite{FH}.\\
Next, the real forms of each of these Lie algebras can be 
classified by identifying the generators which also span a real algebra
(i.e. which admit real structure constants). 
In particular, every simple algebra has a compact form, the real 
algebra over which the Killing form is negative definite. The corresponding
Lie group is compact.  
All the real forms are classified and are described for example in \cite{G}.

\subsection{Lie groups from Lie algebras}
As we have seen in the previous sections, from a Lie group it is easy to
obtain the associated Lie algebra by simple differentiation.
Less trivial is the issue of recovering the group from the algebra.
This is indeed the main argument of the remaining sections.
Here we are simply going to recall some properties of a key tool, the
exponential map:
$$
\exp: Lie(G)\longrightarrow G;\ X\mapsto g_X (1)
$$
where $g_X (t)$ is the integral curve on $G$ associated to the left invariant vector field $X$, with $g_X(0)=e$.
Its main properties are
\begin{itemize}
\item $\exp(0)=e$;
\item $\exp(X+Y)=\exp(X)\exp(Y)$ if $[X,Y]=0$;
\item $\exp$ is differentiable and
$$
d\exp_0 : T_0 Lie(G)\longrightarrow T_e G
$$
realizes the natural isomorphism between $Lie(G)$ and $T_eG$;
\item $\exp$ is a local diffeomorphism between an open neighborhood of $0\in Lie(G)$ and an open neighborhood of $e\in G$.
\end{itemize}
In general the exponential map is not surjective, however it generates
the whole group by starting from the algebra. For matrix groups
it is easy to show that
$$\exp(X)=e^X:=\sum_{n=0}^\infty \frac 1{n!} X^n.$$
As we are going to work with finite representations, this is our case.
Given a matrix realization of the group in a suitable parametrization
$g(x_1,\ldots,x_n)$, the expression for the Cartan 1-form is:
\eq
J=g^{-1} dg =\sum_i J^i \tau_i,
\feq
where $\{\tau_i\}$ is a basis for the Lie algebra. In physics the 
1-forms $J^i$ are also often called the left-invariant currents. 
They will play a central role in our construction.\\
The main problem is now to find suitable parameterizations of the group,
which on the one hand should be able to capture the whole group, but
on the other hand should still remain manageable from a practical point 
of view, i.e. suitable for concrete physical applications. This means that
we need not only to explicitly individuate the elements of the group,
but also to specify the complete range for the parameters, and to compute 
explicitly the significant quantities such as for example the left 
invariant currents, the invariant measure and the Killing form.

\section{CONSTRUCTION OF COMPACT LIE GROUPS}

\subsection{A toy model}\label{sec:toy}
We start by illustrating the main ideas of our strategy in the simplest 
possible example, the construction of the $SU(2)$ group, the set of all 
unitary matrices with unitary determinant.
The associated Lie algebra $\mathfrak{su}(2)$ is generated by the Pauli 
matrices
\eqn
\sigma_1 =\left( \begin{array}{cc} 0 & 1 \\ 1 & 0 \end{array} \right) \ ,
\qquad
\sigma_2 =\left( \begin{array}{cc} 0 & -i \\ i & 0 \end{array} \right)\ ,\qquad
\sigma_3 =\left( \begin{array}{cc} 1 & 0 \\ 0 & -1 \end{array} \right) \ ,
\feqn
which, after multiplication by $i$, indeed constitute a basis for the
space of $2\times 2$ anti Hermitian matrices. It is a well known fact 
that the generic element of $SU(2)$ can be expressed in the form:
\eqn
g= e^{i\phi \frac {\sigma_3}2} e^{i\theta \frac{\sigma_1}2} e^{i\psi \frac{\sigma_3}2}  \ ,  \label{su2}
\feqn
where $\phi \in [0, 2\pi]$, $\theta \in [0, \pi]$, $\psi \in [0,4\pi]$ are
called the {\it Euler angles} for $SU(2)$. Let us first recall the definition of the Euler angles traditionally used in classical
mechanics to describe the motion of a spin. Choose a Cartesian frame 
$(x,y,z)$ and model the spin as a rod of length $L$, with an end
fixed in the origin and the other one in the starting position 
$\vec L\equiv(0,0,L)$. The top of the spin can be moved to a generic 
position in the following way:
\begin{itemize}
\item First, we rotate the system by an angle $\alpha$ around the $z$ axis.
Accordingly, the $x$ axis will be rotated by $\alpha$ to a new axis $x'$ 
in the $x-y$ plane, and similarly for the $y$ axis.
\item Then we rotate the system by an angle $\beta$ around the axis $x'$.
The $z$ axis will be rotated by $\beta$ to a new axis $z''$
in the $y'-z$ plane.
\item Finally, we rotate the system by an angle $\gamma$ around the $z''$ 
axis.
\end{itemize}
Essentially, these movements represent the inclination of the spin with respect to the vertical axis, the rotation around the vertical axis and
the rotation around its proper axis. To describe these operations mathematically, we notice that a rotation  $R_{\hat n} (\theta)$ by an angle 
$\theta$ around an (oriented) axis specified by a unit vector 
$\hat n \equiv (n_x, n_y, n_z)$ can be written as:
\begin{eqnarray}
R_{\hat n} (\theta)=e^{\theta (n_x \tau_1+n_y \tau_2+n_z \tau_3)},
\end{eqnarray}
where
\begin{eqnarray}
\tau_1 =\left( \begin{array}{ccc} 0 & 0 & 0 \\ 0 & 0 & 1 \\  0 & -1 & 0 \end{array} \right) \ ,
\qquad
\tau_2 =\left( \begin{array}{ccc} 0 & 0 & 1 \\ 0 & 0 & 0 \\  -1 & 0 & 0 \end{array} \right) \ , \qquad
\tau_3 =\left( \begin{array}{ccc} 0 & 1 & 0 \\ -1 & 0 & 0 \\  0 & 0 & 0 \end{array} \right)
\end{eqnarray}
are the generators of the infinitesimal rotations. Thus, the generic final position of the top of the spin will be:
\begin{eqnarray}
\vec L' =e^{\gamma \tau_3''} e^{\beta \tau_1'} e^{\alpha \tau_3},
\label{rot}
\end{eqnarray}
where $\tau'_1$ and $\tau''_3$ are the generators of the rotations around 
the $x'$ and $z''$ axis respectively:
\begin{eqnarray}
&& \tau'_1=\cos\alpha \tau_1 +\sin \alpha \tau_2 =e^{\alpha \tau_3} \tau_1 e^{-\alpha \tau_3},\\
&& \tau''_3=\cos\beta \tau_3 -\sin \beta \tau'_2=e^{\beta \tau'_1} \tau_3 e^{-\beta \tau'_1}.
\end{eqnarray}
By remembering that:
$$
e^{e^A B e^{-A}}=e^A e^B e^{-A}
$$
and substituting this in (\ref{rot}), we find
\begin{eqnarray}
\vec L'=e^{\alpha \tau_3} e^{\beta \tau_1} e^{\gamma \tau_3} \vec L.
\end{eqnarray}
{From} this construction it is clear that we can set the range of the Euler
angles to be for example
$\alpha,\gamma\in[0,2\pi],\ \beta\in [0,\pi]$.
This is very similar to (\ref{su2}) and it is tempting to identify 
$\phi, \theta$ and $\psi$ with $\alpha, \beta$ and $\gamma$, respectively.
However, for $SU(2)$ we have $\psi\in [0,4\pi]$ and not $[0,2\pi]$,
which is a consequence of the fact that $SU(2)$ is a double cover
of $SO(3)$ and provides a spin $\frac 12$ representation.

\noindent Let us now look at the structure of the construction (\ref{su2}). We have identified a maximal subgroup 
$U(1)[\phi]=e^{i\phi \frac{\sigma_3}2}$.
Its Lie algebra is obviously a subalgebra of $\mathfrak{su}(2)$. Then we 
have added a second generator $\tau_1$ which does not belong to the
subalgebra, and after observing that all the remaining generators can be 
obtained by commuting $\tau_1$ with the subalgebra, we have acted on 
$e^{i\theta \frac{\sigma_1}2}$
with the subgroup both from the left and from the right:
\begin{eqnarray}
g=U(1)[\phi] e^{i\theta \frac{\sigma_1}2} U(1)[\psi].
\end{eqnarray}
This provides the structure of the generic element of the group, but
more information is still needed in order to determine the minimal 
range for the parameters. \\
For completeness, let us look at the geometric properties of the group and
use them to identify the parameters. It is known that the group $SU(2)$ is 
geometrically equivalent to a three-sphere $S^3$, and admits a Hopf 
fibration structure with fiber $S^1$ over the base 
$S^2 \simeq \mathbb{CP}^1$. To see this, note that by definition the generic
$U(2)$ element can be written in the form
$$
g=\left( \begin{array}{cc} u_1 & w_1 \\ u_2 & w_2 \end{array} \right)
$$
where
$$
\vec u=\left( \begin{array}{c} u_1  \\ u_2  \end{array} \right), \qquad\ \vec w=\left( \begin{array}{c} w_1  \\ w_2  \end{array} \right)
$$
determine an orthonormal basis for $\mathbb{C}^2$. After imposing the condition $\det g=1$ we find that it becomes
$$
g=\left( \begin{array}{cc} u_1 & -u_1^* \\ u_2 & u_2^* \end{array} \right)
$$
where $|u_1|^2+|u_2|^2=1$. Setting $u_1=x+iy$ and $u_2=t+iz$ we see the
correspondence with $S^3$. As we have remarked in the previous section,
since $SU(2)$ is a real compact form, it is naturally endowed with an
invariant metric given by the Killing product. Suitably normalized, this is
$ds^2=-\frac 12 \Tr (g^{-1} dg \otimes g^{-1}dg)$, so that we find
\eq
ds^2=\frac 12 \Tr (dg^\dagger \otimes dg)=\left. (dx^2+dy^2+dt^2+dz^2)\right|_{x^2+y^2+t^2+z^2=1} \label{round}
\feq
which is the usual round metric on the sphere $S^3$. Therefore, to
determine the ranges for the parameters in (\ref{su2}), we can compute
the associated metric, identify it with the round metric and choose the 
range in such a way that it covers the whole $S^3$.
{From} (\ref{su2}) we get:
\eq
ds^2=\frac 14 (d\phi^2 +d\theta^2 +d\psi^2 +2\cos \theta d\phi d\psi) \ ,
\feq
which can be obtained from (\ref{round}) by setting
\eq
u_1= \cos \frac \theta2 e^{\frac i2\epsilon_1(\phi+\psi)+i\alpha_1}, \qquad\ u_2= \sin \frac \theta2 e^{\frac i2\epsilon_2(\phi-\psi)+i\alpha_2} \ ,
\feq
where $\epsilon_i$ are signs and $\phi_i$ constant phases. We do not need to determine these quantities to find the ranges. Indeed, for any
fixed value of these parameters, to cover $S^3$ we need to take
\eq
\frac 12 (\phi+\psi)\in [0,2\pi], \quad\ \frac 12 (\phi-\psi)\in [0,2\pi], \quad\ \theta \in [0,\pi]
\feq
which are equivalent to the ones we announced after (\ref{su2}).\\
In this very simple case all the phases and signs can be determined by
observing that:
\eq
e^{i\phi \frac {\sigma_3}2} e^{i\theta \frac{\sigma_1}2} e^{i\psi \frac{\sigma_3}2}=
\left( \begin{array}{cc} e^{\frac i2 (\phi+\psi)} \cos \frac \theta2 & ie^{\frac i2 (\phi-\psi)} \sin \frac \theta2 \\
ie^{-\frac i2 (\phi-\psi)} \sin \frac \theta2 & e^{-\frac i2 (\phi+\psi)} \cos \frac \theta2 \end{array} \right) \ ,
\feq
which yields $\epsilon_1=1$, $\epsilon_2=-1$, $\phi_1=0$ and $\phi_2=-\frac \pi2$.\\
This is a geometric technique to determine the ranges, but there is 
another tehnique which is much simpler for higher dimensional groups $G$.
It consists in studying the maximal subgroup $U$ of $G$ and the 
quotient $G/U$ separately. In our case we can take $U=U(1)[\psi]$. This is 
a circle with metric $\frac 14 d\psi^2$.
The range of $\psi$ must be a period covering the whole circle, and, being
$$
U(1)[\psi]=e^{i\psi \frac{\sigma_3}2}=\left(
\begin{array}{cc} 
e^{\frac i2\psi} & 0 \\
0 & e^{-\frac i2\psi}
\end{array} \right) \ ,
$$
we can take $\psi\in [0,4\pi]$.
The points of the quotient are parameterized by
\eq\label{H}
H(\phi,\theta)=e^{i\phi \frac {\sigma_3}2} e^{i\theta \frac{\sigma_1}2}=
\left( \begin{array}{cc} e^{\frac i2 \phi} \cos \frac \theta2 & ie^{\frac i2 \phi} \sin \frac \theta2 \\
ie^{-\frac i2 \phi} \sin \frac \theta2 & e^{-\frac i2 \phi} \cos \frac \theta2 \end{array} \right)
\feq
with a residual action of $U(1)[\psi]$ on the right. For example, 
we see that in the quotient $H(\phi,0)$ degenerates to a single point 
and similarly for $H(\phi,\pi)$, because
\begin{eqnarray*}
H(\phi,0)=\left( \begin{array}{cc} e^{\frac i2 \phi} & 0 \\ 0 & e^{-\frac i2 \phi}  \end{array} \right)
=\left( \begin{array}{cc} 1 & 0 \\ 0 & 1  \end{array} \right)
\left( \begin{array}{cc} e^{\frac i2 \phi} & 0 \\ 0 & e^{-\frac i2 \phi}  \end{array} \right)\sim
\left( \begin{array}{cc} 1 & 0 \\ 0 & 1  \end{array} \right)\\
H(\phi,\pi)=\left( \begin{array}{cc} 0 & ie^{\frac i2 \phi} \\ ie^{-\frac i2 \phi} & 0  \end{array} \right)
=\left( \begin{array}{cc} 0 & i \\ i & 0  \end{array} \right)
\left( \begin{array}{cc} 0 & e^{-\frac i2 \phi} \\ e^{\frac i2 \phi} & 0 \end{array} \right)\sim
\left( \begin{array}{cc} 0 & i \\ i & 0  \end{array} \right).
\end{eqnarray*}
Indeed, we can take for the quotient the representative
\eq
H(\phi,\pi)U(1)[-\phi]=\left( \begin{array}{cc} \cos \frac \theta2 & ie^{i \phi} \sin \frac \theta2 \\
ie^{-i \phi} \sin \frac \theta2 & \cos \frac \theta2 \end{array} \right)
\feq
so that as $\phi$ and $\theta$ vary within their ranges, this traces a two
dimensional semi sphere $x\geq 0$ in the $(x,0,t,z)$ space. However,
the equator $x=0$ is contracted to a point and the semi sphere reduces to a
sphere $S^2$ of radius $1/2$. This is the celebrated Hopf fibration.
To see this we can compute the metric on the quotient. This is not simply
$$
ds_H^2= -\frac 12 \Tr (H^{-1} dH \otimes H^{-1} dH) \ ,
$$
because $J_H=H^{-1} dH$ is not cotangent to the quotient, having a
component which is cotangent to the fiber. Using (\ref{H}) we have, indeed:
\eq
J_H=\frac i2 (d\theta \sigma_1+ \sin \theta d\phi \sigma_2 +\cos \theta d\phi \sigma_3).
\feq
However, we can simply project out the component along the fiber, i.e.
the part spanned by $\sigma_3$, so that
\eq
\tilde J_H:= \frac i2 (d\theta \sigma_1+ \sin \theta d\phi \sigma_2)
\feq
and we can recover the metric of a two-sphere of radius $\frac 12$:
\eq\label{base}
ds_H^2=-\frac 12 \Tr \tilde J_H \otimes \tilde J_H= \frac 14 \left[ d\theta^2 +\sin^2 \theta d\phi^2 \right] \ .
\feq
Changing to the complex coordinate $z=\tan \psi e^{i\phi}$ and its 
complex conjugate, this metric reduces to the standard Fubini-Study metric
for $\mathbb{CP}^1$. However, in general we can assume the ranges of the 
parameters for the quotient space to be unknown. Then, they can be deduced 
from (\ref{base}) as follows: the metric becomes degenerate at $\theta=0,
\pi$. This is because fixing $\theta$ and varying $\phi$ we obtain a circle
with radius $\frac 12 \sin \theta$. Therefore, we have to restrict 
$\theta$ to $[0,\pi]$. But we don't have such a constraint on $\phi$
and in principle it could vary with a period which we know to be $4\pi$ 
as for $\psi$. However, this is not the right period for the quotient.
Indeed, note that $-I=(^{-1}_{\ 0} {}^{\ 0}_{-1})\in U(1)[\psi]$ and 
it is in the center of the group, so that:
$$
H(\phi,\theta)\sim H(\phi,\theta)(-I)= -H(\phi,\theta)=H(4\pi-\phi,\theta)
\ .
$$
Therefore, $\phi\sim 4\pi-\phi$, which means that $0\sim 2\pi$, reducing
the period to $\phi\in[0,2\pi]$.

\subsection{The generalized Euler construction.}\label{general}
Let us now generalize the previously described construction to the compact form of a generic finite dimensional simple Lie group $G$, $n={\rm dim} G$.
In this case, our construction is not unique but is related to the choice of a maximal subgroup $H$. Because $G$ is compact, the Killing product
defines a scalar product $(|)$ on $\mathfrak{g}=Lie(G)$ and it is convenient to choose an orthonormal basis $\{\tau_i \}_{i=1}^n$ of $\mathfrak{g}$.
In particular, let us assume that the first $k:= {\rm dim} H$ generators
are a basis for $\mathfrak{h}=Lie (H)$ and let us call $\cP$ the subspace 
spanned by the remaining generators. Note that $[\cHH,\cP]\subset \cP$.
Indeed, orthogonality and $ad$-invariance imply
$$
([p,h]|h')=(p|[h,h'])=0
$$
for any $p\in \cP$ and $h,h'\in \cHH$. This means that $G/H$ is reductive. From this, it follows that any $g\in G$ can be written in the form
\eqn
g=\exp a \exp b \ , \quad \ a\in \cP\ , \ b\in \cHH \ .
\feqn
For compact simple Lie groups such a parametrization is surjective, a proof can be found in \cite{BCCS1}.\\
Now, let's suppose we have an explicit parametrization for $H$, which
is obviously a generalized Euler parametrization obtained inductively by
choosing a maximal subgroup $H'$ of $H$ and proceeding in the same way.
This means that we can use the parametrization to give an expression 
for $\exp b$. Now we would like to improve the expression for $\exp a$. 
To this purpose we can look for a subset of linearly free elements
$\tau_1,\ldots,\tau_l\in \cP$ with the following properties:
\begin{itemize}
\item if $V$ is the linear subspace generated by $\tau_i$, $i=1,\ldots,l$, then $\cP=Ad_H (V)$, that is, the whole $\cP$ is generated from $V$ through
the adjoint action of $H$;
\item $V$ is minimal, in the sense that it does not contain any proper subspaces with the previous property.
\end{itemize}
Because of simplicity, it is not hard to show that such a subspace $V$
of $\cP$ always exists.
Therefore, the general element $g$ of $G$ can be written in the form:
\eqn
g= \exp (\tilde b) \exp (v) \exp (b)\ , \quad b,\tilde b \in \cHH\ ,\ v\in V \ .
\feqn
This parametrization is obviously redundant, since in general it depends
on $2k+l\geq n$ parameters. The point is that not the whole of $H$ is needed
to generate $V$ by the adjoint action, because $H$ contains some
subgroup $H_o$ which generates the automorphisms of $V$:
\eqn
Ad_{H_o} : V\longrightarrow V \ .
\feqn
Then $H_o$ must be $r$-dimensional, where $r=2k+l-n$ is the redundancy, and 
the generalized Euler decomposition with respect to $H$ finally takes the form
\eq
G=B \exp (V) H \ , \label{eulero}
\feq
where $B:=H/H_o$. We have seen that even for the simplest case of $SU(2)$ 
the automorphism group $H_o$ is not trivial (even though it acts trivially on
$V$) and it coincides with $\mathbb {Z}_2$.

\subsection{Determination of the range of parameters}
The symbolic expression (\ref{eulero}) means that the generic element of $G$ can be written in the form
\eqn
g=b e^v h, \qquad b\in B, \ v\in V, \ h\in H,
\feqn
where $h$, $v$ and $b$ are function of $k, l$ and $n-l-k$ parameters respectively.
Locally, they define a coordinatization for the group.
However, being the parametrization surjective, the parameters can be chosen 
in such a way as to cover the whole group. However, because in general the group
is a non trivial manifold, a surjective parametrization cannot in general 
be injective. A good choice for the range of the parameters is to pick a maximal
open subset on which the parametrization is injective, so that its closure covers
the whole group. We will call this closure {\it
the range of parameters}. In general the determination of the range is a highly
non trivial task. The aim here is to discuss two practical methods to do this.

\subsubsection{Geometric identification}
Once the parametrization $g[\vec x]$ is given, it can be used to describe the
geometry of the group or of its quotient with the maximal subgroup. If such a
geometry is already known by some other means, this information can be applied
to determine the range of the parameters.\\
The metric on the group can be computed by starting from the Killing metric
and the Cartan 1-form. The parametrization provides a local
coordinatization, which in turns yields a local expression for the 
Cartan 1-form:
\eqn
J = g^{-1} \frac {\partial g}{\partial x^J} dx^J =J^i \tau_i \ ,
\feqn
where $\tau_i$, $i=1,2,\ldots,n$ is a basis for the Lie algebra. This defines the structure constants
${f_{ij}}^k$ so that the Killing metric has components:
\eqn
K_{ij}=-k{f_{il}}^m {f_{jm}}^l \ ,
\feqn
where $k$ is some normalization constant. As we are working with a compact form, the metric is positive definite when $k$ is positive.
We choose the basis and $k$ in such a way that $K_{ij}=\delta_{ij}$. The metric induced on the manifold is then
\eqn
ds^2 =g_{ij} dx^i \otimes dx^j = J^l \otimes J^m \delta_{lm} \ .
\feqn
In other words, the 1-forms $J^l$ represent the vielbein one forms on the group.
In particular they can be used to compute the invariant volume $n$-form
\eqn
\omega=J^1 \wedge \ldots \wedge J^{n}= \det (\underline J) dx^1 \wedge \ldots \wedge dx^{n}
\label{volform}
\feqn
and the corresponding Haar measure
\eq
d\mu =|\det (\underline J)| \prod_{I=1}^{n} dx^I \ . \label{measure}
\feq
{From} our parametrization (\ref{eulero}), we can write the general element $g\in G$ in the form
\eqn
g(x_1,\ldots,x_s, y_1,\ldots, x_m)=p(x_1,\ldots,x_s) h(y_1, \ldots, y_m)
\feqn
where $h\in H$, $p\in B\exp (V)$, $m={\rm dim} H$ and $m+s=n$.
We can assume for simplicity that $\{\tau_a\}$, $a=s+1,\ldots,n$ generates $H$.
Notice that only $H$ is a subgroup, so that $J_h \equiv h^{-1} dh \in Lie(H)$,
whereas in general $J_p\equiv p^{-1} dp \in Lie(G)$.
However, for the subgroup, instead of the left-invariant form we prefer to use the right-invariant form $\tilde J_h \equiv dh\ h^{-1}$.
In this way, setting
\eqn
J_p = \sum_{i=1}^n J_p^i \tau_i \ , \qquad \tilde J_h = \sum_{i=s+1}^{n} \tilde J_h^i \tau_i \ ,
\feqn
and using orthonormality, after a simple calculation we get
\eq
ds^2 = \sum_{i=s+1}^n \left( J_p^i +\tilde J_h^i \right)^2 +\sum_{i=1}^s \left( J_p^i \right)^2 . \label{decompmetr}
\feq
{From} this expression for the metric we can read the structure of the fibration with fiber $H$ over $G/H$. Indeed, the forms
$J_p^i +\tilde J_h^i$, $i=s+1,\ldots,n$ lie on the fiber, whereas $J_p^a$, $a=1,\ldots,s$ are orthogonal to the fiber.
This means that
\begin{equation}\label{quoz}
d\sigma^2 =\sum_{i=1}^s \left( J_p^i \right)^2
\end{equation}
defines the metric on the quotient space $G/H$, as defined
by the $s$-dimensional vielbein $\hat J_p$ obtained from $J_p$ by projecting out the components along the fiber.
At a practical level, this decomposition not only allows us to perform the
computation of the metric on the quotient space, but it also greatly
simplifies the explicit calculation of the metric on the whole group.\\
To conclude, this method can be used to provide an explicit characterization
of the geometry of the group and of its quotients and if some of the geometrical
structure of the group and/or of its fibration is known by any other means, by 
comparison we can determine the range of the parameters. We are going to see an 
explicit example of this procedure later.

\subsubsection{A topological method}\label{sec:topo}
In general, however, the theoretical information about the group is not
sufficient to determine the explicit range for the parameters.
In this case, we need to introduce an alternative method which requires a 
minimal amount of information to work. Fortunately, such a method,
which we called topological, is provided by a powerful theorem due to
I.G. Macdonald which describes a simple way to compute the total volume of a 
compact connected simple Lie group. Let $\mathfrak{c} \subset Lie(G)$ be a
Cartan subalgebra, and $\mathfrak{c}_{\mathbb{Z}}$ the integer lattice generated
in $\mathfrak{c}$ by a choice of simple roots (the root lattice).
Then, the first geometrical ingredient is the torus $T :=\mathfrak{c}/\mathfrak{c}_{\mathbb{Z}}$, whose dimension is $r={\rm rank} Lie(G)$.
The second ingredient is a well known result due to Hopf \cite{H}: the rational
homology of $G$ is equal to the rational homology of a product
of odd-dimensional spheres
$$
H_* (G,\mathbb{Q}) \simeq H_* \left(\prod_{i=1}^k (S^{2i+1})^{r_i},\mathbb{Q} \right) ,
$$
where $r_i$ is the number of times the given sphere appears, and $r_1+\ldots+r_k=r$.
The result of Macdonald \cite{M} can then be stated as follows:\\
If we assign a Lebesgue measure $\mu$ on a compact simple Lie group $G$ by means of an Euclidean scalar product $\langle \ , \rangle$ on
$\mathfrak{g}=Lie(G)$,
then the measure of the whole group is
\begin{equation}\label{macdonald}
\mu(G)=\mu_o({T}) \cdot \prod_{i=1}^k Vol(S^{2i+1})^{r_i} \cdot \prod_{\alpha \in R(\mathfrak{g})} \frac 2{|\alpha|} \ ,
\end{equation}
where $R(\mathfrak{g})$ is the set of non vanishing roots, $\mu_o$ is the Lebesgue measure on $\mathfrak{g}$ induced by the scalar product
and $Vol(S^{2i+1}) =2\pi^{i+1}/i!$ is the volume of the unit sphere $S^{2i+1}$.\\
On the other side, we can in principle compute the measure of the whole group, induced by the Killing scalar product, by using
(\ref{measure}) integrated over the range of the parameters. Using (\ref{decompmetr}) we get
\eq
d\mu =|\det (\underline J_p)| |\det (\underline {\tilde J}_h)| \prod_{I=1}^{n} dx^I \ . \label{measure1}
\feq
Now, let's assume we have a good parametrization, which means that the one 
parameter subgroups spanned by the orbits $\exp (t\tau_i)$,
where $\{\tau_i\}$ is the basis we fixed for the Lie algebra, are subgroups
embedded in $G$\footnote{This can be accomplished for any simple group.}.
Then, such orbits are compact (for a compact group) and $\exp (t\tau_i)$ is
periodic in $T$. The point is that if we choose correctly the range
${\mathcal R}$ for the parameters, then
\begin{equation}
\mu(G)=\int_{\mathcal R} d\mu.
\end{equation}
A ``suitable'' range means that it covers each point of $G$ exactly once, up to a
subset of vanishing measure.
Let us look at the measure weight $f:=|\det (\underline J)|$. In general it will depend explicitly on the parameter but not on all the parameters.
For each parameter which does not appears in $f$ we choose its period as a range. Let us call $\bar {\mathcal R}$ the range for the remaining
parameters. Then, $\bar {\mathcal R}$ has a boundary defined by $f=0$. This equation in general provides a splitting of the space into infinitely
many fundamental regions, which, however, turn out to be all equivalent
for our purposes. With such a choice ${\mathcal R}_o$ for the range, we are
sure that its image under our parametrization map 
$g : {\mathcal R}_o \rightarrow G$ describes a closed $n$-dimensional 
variety on $G$. As $G$ is connected, $g({\mathcal R}_o)$ has to cover $G$
an integer number $m$ of times
\eq
m=\frac 1{\mu (G)} \int_{{\mathcal R}_o} d\mu.
\feq
If $m>1$ it means that it exist an automorphism group $\Gamma: {\mathcal R}_o\rightarrow {\mathcal R}_o$ of order $m$, such that
$g(\Gamma x)=g(x)$. In this case we restrict the range to
\eq
{\mathcal R}={\mathcal R}_o/\Gamma.
\feq
This is, indeed, what we have done at the end of section \ref{sec:toy}.
Let us now illustrate our procedure for some exceptional examples.

\section{GENERALIZED EULER ANGLES FOR \boldmath{$G_2$}}\label{sec:g2}

\subsection{The Lie algebra}
\begin{figure}[h]\label{fig1}
\begin{center}
\includegraphics[angle=90,scale=0.35]{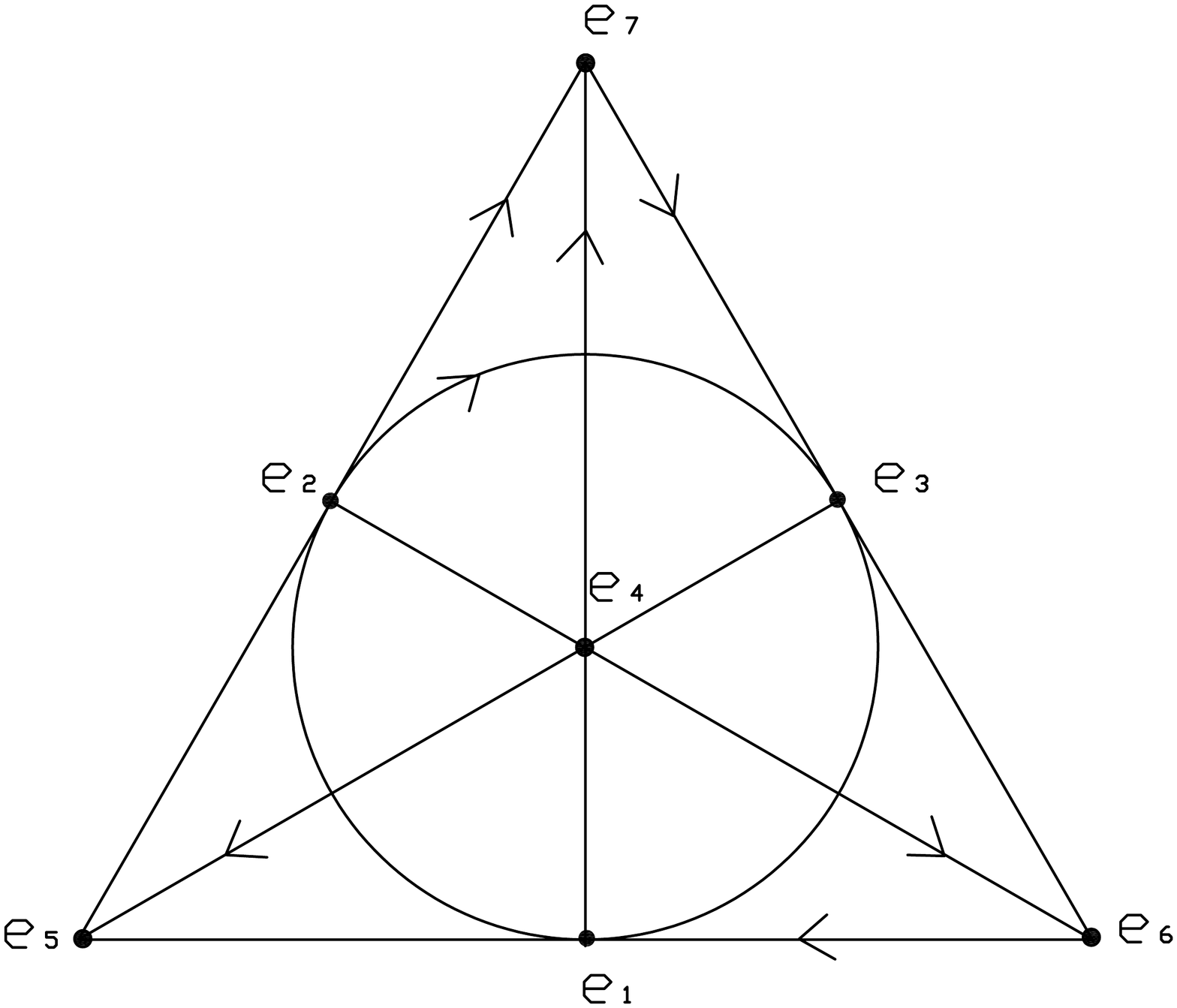}
\caption{\em The Fano diagram.}
\end{center}
\end{figure}

The exceptional Lie group $G_2$ can be realized as the automorphism group of the
octonionic algebra \cite{B,A}. Instead of providing a theoretical proof
of this fact, we explicitly construct such a group starting for its Lie algebra.\\
The octonionic algebra $\mathbb O$ is the eight dimensional real vector field generated by a real unit $e_0\equiv 1$ and seven imaginary units
$e_i$, $i=1,\ldots,7$. It is endowed with a distributive but non associative 
product described by the relations:
\begin{eqnarray*}
&& e_0 \cdot a= a \cdot e_0 =a \quad\ \forall a\in \mathbb{O},\\
&& e_i^2=-e_0, \quad e_i\cdot e_j =- e_j\cdot e_i, \qquad 1\leq i<j \leq 7
\end{eqnarray*}
and the Fano diagram, see fig. \ref{fig1}.
Each oriented line can be thought as an oriented circle, on which three distinct
imaginary roots $e_i, e_j, e_k$ are lying, the products of which are
$e_i\cdot e_j=\pm e_k$. Here the sign is positive if and only if the triple
$\{ e_i, e_j, e_k\}$ follows the orientation of the arrow. For example,
$e_1 \cdot e_3=-e_2$ and $e_1\cdot e_2=e_3$. Notice that each circle generates a 
quaternionic subalgebra.
An automorphism of the algebra is an invertible linear map:
$$
A: \mathbb{O} \longrightarrow \mathbb{O}
$$
satisfying
$$
A(a\cdot b)=A(a)\cdot A(b), \qquad a,b\in \mathbb{O}.
$$
The set of all automorphisms is a group with respect to the composition product, and is indeed a Lie group. From this it follows immediately that
its Lie algebra is the set of derivations $\mathfrak {D}(\mathbb O)$, the linear operators
\eq
B: \mathbb{O} \longrightarrow \mathbb{O}
\feq
satisfying
\eq\label{derivation}
B(a\cdot b)=B(a)\cdot b+ a\cdot B(b), \qquad a,b\in \mathbb{O},
\feq
and with the commutator as Lie product. Note that $B(1)=0$ for all
$B\in \mathfrak {D}(\mathbb O)$, so that we can look for a matrix representation
of $\mathfrak {D}(\mathbb O)$ on the real space spanned by the imaginary units.
This will give the smallest fundamental representation of $G_2$, the 
{\boldmath{$7$}} representation. Imposing the condition (\ref{derivation}),
with the help of a computer, we find a set of $14$ linearly independent matrices:
$$
C_1 =\left(
\begin{array}{ccccccc}
0 & 0 & 0 & 0 & 0 & 0 & 0 \\
0 & 0 & 0 & 0 & 0 & 0 & 0 \\
0 & 0 & 0 & 0 & 0 & 0 & 0 \\
0 & 0 & 0 & 0 & 0 & 0 & -1 \\
0 & 0 & 0 & 0 & 0 & -1 & 0 \\
0 & 0 & 0 & 0 & 1 & 0 & 0 \\
0 & 0 & 0 & 1 & 0 & 0 & 0
\end{array}
\right),
\qquad
 C_2 =\left(
\begin{array}{ccccccc}
0 & 0 & 0 & 0 & 0 & 0 & 0 \\
0 & 0 & 0 & 0 & 0 & 0 & 0 \\
0 & 0 & 0 & 0 & 0 & 0 & 0 \\
0 & 0 & 0 & 0 & 0 & 1 & 0 \\
0 & 0 & 0 & 0 & 0 & 0 & -1 \\
0 & 0 & 0 & -1 & 0 & 0 & 0 \\
0 & 0 & 0 & 0 & 1 & 0 & 0
\end{array}
\right),
$$
$$
C_3 =\left(
\begin{array}{ccccccc}
0 & 0 & 0 & 0 & 0 & 0 & 0 \\
0 & 0 & 0 & 0 & 0 & 0 & 0 \\
0 & 0 & 0 & 0 & 0 & 0 & 0 \\
0 & 0 & 0 & 0 & -1 & 0 & 0 \\
0 & 0 & 0 & 1 & 0 & 0 & 0 \\
0 & 0 & 0 & 0 & 0 & 0 & -1 \\
0 & 0 & 0 & 0 & 0 & 1 & 0
\end{array}
\right),
\qquad
C_4 =\left(
\begin{array}{ccccccc}
0 & 0 & 0 & 0 & 0 & 0 & 0 \\
0 & 0 & 0 & 0 & 0 & 0 & 1 \\
0 & 0 & 0 & 0 & 0 & 1 & 0 \\
0 & 0 & 0 & 0 & 0 & 0 & 0 \\
0 & 0 & 0 & 0 & 0 & 0 & 0 \\
0 & 0 & -1 & 0 & 0 & 0 & 0 \\
0 & -1 & 0 & 0 & 0 & 0 & 0
\end{array}
\right),
$$
$$
C_5 =\left(
\begin{array}{ccccccc}
0 & 0 & 0 & 0 & 0 & 0 & 0 \\
0 & 0 & 0 & 0 & 0 & -1 & 0 \\
0 & 0 & 0 & 0 & 0 & 0 & 1 \\
0 & 0 & 0 & 0 & 0 & 0 & 0 \\
0 & 0 & 0 & 0 & 0 & 0 & 0 \\
0 & 1 & 0 & 0 & 0 & 0 & 0 \\
0 & 0 & -1 & 0 & 0 & 0 & 0
\end{array}
\right),
\qquad
C_6 =\left(
\begin{array}{ccccccc}
0 & 0 & 0 & 0 & 0 & 0 & 0 \\
0 & 0 & 0 & 0 & 1 & 0 & 0 \\
0 & 0 & 0 & -1 & 0 & 0 & 0 \\
0 & 0 & 1 & 0 & 0 & 0 & 0 \\
0 & -1 & 0 & 0 & 0 & 0 & 0 \\
0 & 0 & 0 & 0 & 0 & 0 & 0 \\
0 & 0 & 0 & 0 & 0 & 0 & 0
\end{array}
\right),
$$
$$
C_7 =\left(
\begin{array}{ccccccc}
0 & 0 & 0 & 0 & 0 & 0 & 0 \\
0 & 0 & 0 & -1 & 0 & 0 & 0 \\
0 & 0 & 0 & 0 & -1 & 0 & 0 \\
0 & 1 & 0 & 0 & 0 & 0 & 0 \\
0 & 0 & 1 & 0 & 0 & 0 & 0 \\
0 & 0 & 0 & 0 & 0 & 0 & 0 \\
0 & 0 & 0 & 0 & 0 & 0 & 0
\end{array}
\right),
\qquad
C_8 =\frac 1{\sqrt 3} \left(
\begin{array}{ccccccc}
0 & 0 & 0 & 0 & 0 & 0 & 0 \\
0 & 0 & -2 & 0 & 0 & 0 & 0 \\
0 & 2 & 0 & 0 & 0 & 0 & 0 \\
0 & 0 & 0 & 0 & 1 & 0 & 0 \\
0 & 0 & 0 & -1 & 0 & 0 & 0 \\
0 & 0 & 0 & 0 & 0 & 0 & -1 \\
0 & 0 & 0 & 0 & 0 & 1 & 0
\end{array}
\right),
$$
$$
C_9 =\frac 1{\sqrt 3} \left(
\begin{array}{ccccccc}
0 & -2 & 0 & 0 & 0 & 0 & 0 \\
2 & 0 & 0 & 0 & 0 & 0 & 0 \\
0 & 0 & 0 & 0 & 0 & 0 & 0 \\
0 & 0 & 0 & 0 & 0 & 0 & 1 \\
0 & 0 & 0 & 0 & 0 & -1 & 0 \\
0 & 0 & 0 & 0 & 1 & 0 & 0 \\
0 & 0 & 0 & -1 & 0 & 0 & 0
\end{array}
\right),
\qquad
C_{10} =\frac 1{\sqrt 3} \left(
\begin{array}{ccccccc}
0 & 0 & -2 & 0 & 0 & 0 & 0 \\
0 & 0 & 0 & 0 & 0 & 0 & 0 \\
2 & 0 & 0 & 0 & 0 & 0 & 0 \\
0 & 0 & 0 & 0 & 0 & -1 & 0 \\
0 & 0 & 0 & 0 & 0 & 0 & -1 \\
0 & 0 & 0 & 1 & 0 & 0 & 0 \\
0 & 0 & 0 & 0 & 1 & 0 & 0
\end{array}
\right),
$$
$$
 C_{11} = \frac 1{\sqrt 3} \left(
\begin{array}{ccccccc}
0 & 0 & 0 & -2 & 0 & 0 & 0 \\
0 & 0 & 0 & 0 & 0 & 0 & -1 \\
0 & 0 & 0 & 0 & 0 & 1 & 0 \\
2 & 0 & 0 & 0 & 0 & 0 & 0 \\
0 & 0 & 0 & 0 & 0 & 0 & 0 \\
0 & 0 & -1 & 0 & 0 & 0 & 0 \\
0 & 1 & 0 & 0 & 0 & 0 & 0
\end{array}
\right),
\qquad
C_{12} = \frac 1{\sqrt 3} \left(
\begin{array}{ccccccc}
0 & 0 & 0 & 0 & -2 & 0 & 0 \\
0 & 0 & 0 & 0 & 0 & 1 & 0 \\
0 & 0 & 0 & 0 & 0 & 0 & 1 \\
0 & 0 & 0 & 0 & 0 & 0 & 0 \\
2 & 0 & 0 & 0 & 0 & 0 & 0 \\
0 & -1 & 0 & 0 & 0 & 0 & 0 \\
0 & 0 & -1 & 0 & 0 & 0 & 0
\end{array}
\right),
$$
$$
C_{13} = \frac 1{\sqrt 3} \left(
\begin{array}{ccccccc}
0 & 0 & 0 & 0 & 0 & -2 & 0 \\
0 & 0 & 0 & 0 & -1 & 0 & 0 \\
0 & 0 & 0 & -1 & 0 & 0 & 0 \\
0 & 0 & 1 & 0 & 0 & 0 & 0 \\
0 & 1 & 0 & 0 & 0 & 0 & 0 \\
2 & 0 & 0 & 0 & 0 & 0 & 0 \\
0 & 0 & 0 & 0 & 0 & 0 & 0
\end{array}
\right),
\qquad
C_{14} = \frac 1{\sqrt 3} \left(
\begin{array}{ccccccc}
0 & 0 & 0 & 0 & 0 & 0 & -2 \\
0 & 0 & 0 & 1 & 0 & 0 & 0 \\
0 & 0 & 0 & 0 & -1 & 0 & 0 \\
0 & -1 & 0 & 0 & 0 & 0 & 0 \\
0 & 0 & 1 & 0 & 0 & 0 & 0 \\
0 & 0 & 0 & 0 & 0 & 0 & 0 \\
2 & 0 & 0 & 0 & 0 & 0 & 0
\end{array}
\right).
$$
It is easy to check that these matrices do, indeed, define a Lie algebra 
with the commutator product, and that $\mathbb{R}^7$ is irreducible under 
their action, so that they realize an irreducible representation. A rank
two Cartan subalgebra is generated by $C_5,C_{11}$ and in the adjoint
representation it is easy to compute all roots which turn out to coincide
with the roots of $G_2$, as expected (see for example \cite{CCdVOS}).

\subsection{Two Euler parameterizations}
We can now realize two distinct Euler parameterizations for $G_2$, based on different choices of the maximal subgroup $H$. The first
one is based on $H=SU(3)$, \cite{C}, and the second one on $H=SO(4)$, \cite{CCdVOS}.
While for the first one it is possible to apply the geometrical method,
for the second one the topological method is necessary. We are going to call
them the $SU(3)$-Euler parametrization and the $SO(4)$-Euler 
parametrization respectively.
\subsubsection{The \boldmath{$SU(3)$}-Euler parametrization}
Among the automorphisms of the octonions, we can look at the subgroup which
leaves an imaginary unit fixed. This is a subgroup of $G_2$ and will
be contained in the $SO(6)$ group which rotates the remaining six imaginary
units. Indeed, it turns out to be an $SU(3)$ group. We can see it
immediately from our matrices: the first row and column of the first eight
matrices vanish, so that they leave $e_1$ fixed. They generate
a subalgebra, and in the adjoint representation it can be verified that 
the roots match with $SU(3)$. It acts transitively on the subset of 
imaginary units orthogonal to $e_1$, defining a six dimensional sphere
$S^6$, so that $G_2/SU(3)\simeq S^3$. \\
We then choose $\{C_i\}$, $i=1,2,\ldots,8$ as generators for $H$, so that 
$\{C_a\}$, $a=9,\ldots,14$ generate $\mathfrak{p}$.
To identify $V$ (see (\ref{eulero})) we note that $C_9$ generates the whole
$\mathfrak{p}$ under the action of $H=SU(3)$, and, therefore,
$V=\mathbb{R} C_9$.
Finally, note that the subalgebra of $H$ commuting with $C_9$ is the $\mathfrak {su}(2)$ algebra generated by $\{C_i\}$, $i=1,2,3$.
Thus $B=SU(3)/SU(2)$.\\
As a first step we need to construct the $SU(3)$ subgroup $H$. We could proceed in the same way, but as the construction of $SU(3)$ is
well known, we limit ourselves here only to the final result (see 
\cite{BCC,TS,C}):
\eqn\label{subH}
H[x_1, \ldots ,x_8] =e^{x_1 C_3} e^{x_2 C_2}e^{x_3 C_3}
e^{x_4 C_5}e^{\sqrt 3 x_5 C_8}e^{x C_3}e^{x C_2}e^{x_8 C_3} ,
\feqn
with range
\eqn
&& x_1 \in \left[ 0\ , \pi \right] , \qquad
x_2 \in \left[ 0\ , \frac \pi2 \right] , \qquad
x_3 \in \left[ 0\ , \pi \right] , \qquad
x_4 \in \left[ 0\ , \frac \pi2 \right] , \cr
&& x_5 \in \left[ 0\ , 2\pi \right] , \qquad
x_6 \in \left[ 0\ , 2\pi \right] , \qquad
x_7 \in \left[ 0\ , \frac \pi2 \right] , \qquad
x_8 \in \left[ 0\ , \pi \right] .
\feqn
We just want to remark that (\ref{subH}) has the structure of 
(\ref{eulero}) with
\begin{eqnarray}
B=SO(3)=SU(2)/\mathbb{Z}_2, \qquad V=\mathbb{R} C_5, \qquad H=U(2).
\end{eqnarray}
Then, our $SU(3)$-Euler parametrization is
\begin{eqnarray}
g[x_1,\ldots,x_{14}]=
e^{x_1 C_3} e^{x_2 C_2} e^{x_3 C_3} e^{\frac {\sqrt 3}2 x_4 C_8}
e^{x_5 C_5} e^{\frac {\sqrt 3}2 x_6 C_9}
H[x_7 \ , \ldots \ ,x_{14}],
\end{eqnarray}
where we need to determine the range for $x_1,\ldots,x_6$, whereas the remaining parameters have the range of $SU(3)$.
To this aim, we will use the information
$$
G_2/SU(3)\simeq S^3.
$$
From
\eq
p[x_1,\ldots,x_6] = e^{x_1 C_3} e^{x_2 C_2} e^{x_3 C_3} e^{\frac {\sqrt 3}2 x_4 C_8}
e^{x_5 C_5} e^{\frac {\sqrt 3}2 x_6 C_9}
\feq
we can compute $J_p=p^{-1} dp$ and then the metric (\ref{quoz}) induced on the quotient. By a direct computation, we get
\eqn
\frac 43 d\sigma^2 =dx_6^2 +\sin^2 x_6
\left\{ dx_5^2 +\cos^2 x_5 dx_4^2 +\sin^2 x_5 \left[
s_1^2 +s_2^2 +\left( s_3+\frac 12 dx_4 \right)^2 \right] \right\} \label{6metric}
\feqn
where
\eqn
&& s_1 =-\sin (2x_2) \cos (2x_3) dx_1 +\sin(2x_3)dx_2 \cr
&& s_2 =\sin (2x_2) \sin (2x_3) dx_1 +\cos(2x_3)dx_2 \cr
&& s_3 =\cos (2x_2) dx_1 +dx_3.
\feqn
We recognize this as the metric of a round six sphere $S^6$ of radius $\sqrt 3/2$, with coordinates
$(x_6\ , \vec X )$, where $x_6$ is an azimuthal coordinate,
$x \in [0, \pi]$, and $\vec X$ cover a five sphere embedded in $\mathbb C^3$ via
\eqn
&& \vec X =(z_1 \ , z_2 \ , z_3 ) =
\left( \cos x_5 e^{ix_4} \ ,
\sin x_5 \cos x_2
e^{i\left( x_1 +x_3 +\frac {x_4}2\right)} \ ,
\sin x_5 \sin x_2
e^{i\left( x_1 -x_3 -\frac {x_4}2\right)} \right) \ , \cr
&& x_1 \in \left[ 0\ , \pi \right] \ , \qquad
x_2 \in \left[ 0\ , \frac \pi2 \right] \ , \qquad
x_3 \in \left[ 0\ , 2\pi \right] \ , \qquad
x_4 \in \left[ 0\ , 2\pi \right] \ , \qquad
x_5 \in \left[ 0\ , \frac \pi2 \right]  \ . \nonumber
\feqn
Computing the metric $ds^2_{S^5} =|dz_1|^2 +|dz_2|^2 +|dz_3|^2$ in these
coordinates we find
\eqn
\frac 43 d\sigma^2=dx_6^2 +\sin^2 x_6 \left\{ ds^2_{S^5}  \right\} \ .
\feqn
This completes our identification for the range of the parameters.

\subsubsection{The \boldmath{$SO(4)$}-Euler parametrization}
The maximal subgroup $SO(4)$ can be singled out as follows. We know that $1,e_1,e_2, e_3$ generate a quaternionic subalgebra $\mathbb{H}$. We look
at the subgroup $H$ which leaves this subalgebra invariant. This will be generated by block diagonal matrices of the form 
$\{3\times 3\} \times \{4\times 4\}$, which turn out to be the matrices 
$C_i$, $i=1,2,3,8,9,10$. Indeed, $C_1, C_2, C_3$ generate an 
$\mathfrak{su}(2)$ subalgebra, which leaves each element of 
$\mathbb{H}$ invariant.
Let us call this group $SU(2)_I$. Then $C_8, C_{9}, C_{10}$ span a second
$SU(2)$ group ($SU(2)_{II}$), the action of which, when restricted to 
$e_1,e_2,e_3$, generates the automorphisms of $\mathbb{H}$.
Notice that the two subgroups commute. We can now define the surjective
homomorphism:
\begin{eqnarray*}
\phi:  SU(2)_I\times SU(2)_{II} & \longrightarrow & H \\
(a,b) & \mapsto & ab.
\end{eqnarray*}
Observe that ${\rm Ker}\phi $ is the $\mathbb{Z}_2$ subgroup generated by 
the element $(\exp(\pi C_1),\exp(\sqrt 3 \pi C_8))=(z,z)$, with 
$z={\rm diag}\{I_3, -I_4\}$, $I_n$ being the $n\times n$ identity matrix. 
Thus, we can finally obtain $SO(4)$ as:
\eq
H\equiv SU(2)_I\times SU(2)_{II}/\mathbb{Z}_2 =SO(4).
\feq
Its Euler parametrization can be constructed very easily by starting from 
the one for $SU(2)_I$ and $SU(2)_{II}$:
we get
\begin{equation}
H(x_1,\ldots,x_6)=e^{x_1 C_3} e^{x_2 C_2} e^{x_3 C_3} e^{\sqrt 3 x_4 C_8} e^{\sqrt 3 x_5 C_9} e^{\sqrt 3 x_6 C_8},
\end{equation}
where the range is:
\begin{eqnarray}
&& x_1 \in [0,2\pi], \qquad x_{2} \in [0,\pi/2], \qquad x_{3}\in [0,\pi] \cr
&& x_{4} \in [0,\pi], \qquad x_{5} \in [0,\pi/2], \qquad x_{6}\in [0,\pi].
\end{eqnarray}
We also know that $C_{5}, C_{11}$ generate a Cartan subalgebra, not contained in $Lie(H)$. The action of $H$ on this Cartan subalgebra
generates the complement of $Lie(H)$, so that we can take $V=\mathbb{R}C_5\oplus \mathbb{R}C_{11}$. Finally, because of
$$
{\rm dim } B= {\rm dim } G_2 -{\rm dim } H -{\rm dim } V=6 ={\rm dim } H,
$$
we expect for the subgroup $H_o$ of $H$ which commute with $\exp V$ to be a finite group. This means that the $SO(4)$-Euler parametrization
will take the form
\begin{equation}
g[x_1,\ldots,x_{14}]=H(x_1,\ldots,x_6) e^{\sqrt 3 x_7 C_{11} +x_8 C_5} H(x_9,\ldots,x_{14}), \label{so4param}
\end{equation}
where $x_9,\ldots, x_{14}$ will span the whole $SO(4)$, whereas the range
of the first six parameters will be restricted by the action of $H_o$. \\
Before determining $H_o$, we remark that in this case the quotient manifold $M=G_2/SO(4)$ is known to be the eight-dimensional
variety of the quaternionic subalgebras of $\mathbb{O}$. Unfortunately, we cannot use this information as we did before for the $SU(3)$ Euler angles,
because an invariant metric on $M$ (independent from the one we can compute
by group theoretical arguments) is not known, so we have to revert to the 
topological method instead.  \\
Let us now proceed with the determination of $H_o$. This is the subgroup of $7\times 7$ orthogonal matrices $A$ of $SO(4)$,
whose adjoint action leave the Cartan subalgebra invariant:
\eqn
A C_i A^t =C_i, \quad i=5,11.
\feqn
A direct computation shows that it is the finite group $\mathbb{Z}_2 \times \mathbb{Z}_2$ generated by the idempotent
matrices $\sigma$ ($\sigma=\sigma^{-1}$) and $\eta$ ($\eta=\eta^{-1}$)
\eqn
\begin{array}{rcl}
\sigma=\left(
\begin{array}{ccccccc}
1 & 0 & 0 & 0 & 0 & 0 & 0 \\
0 & -1 & 0 & 0 & 0 & 0 & 0 \\
0 & 0 & -1 & 0 & 0 & 0 & 0 \\
0 & 0 & 0 & 1 & 0 & 0 & 0 \\
0 & 0 & 0 & 0 & 1 & 0 & 0 \\
0 & 0 & 0 & 0 & 0 & -1 & 0 \\
0 & 0 & 0 & 0 & 0 & 0 & -1
\end{array}
\right)
\qquad \eta =\left(
\begin{array}{ccccccc}
-1 & 0 & 0 & 0 & 0 & 0 & 0 \\
0 & 0 & 1 & 0 & 0 & 0 & 0 \\
0 & 1 & 0 & 0 & 0 & 0 & 0 \\
0 & 0 & 0 & -1 & 0 & 0 & 0 \\
0 & 0 & 0 & 0 & 1 & 0 & 0 \\
0 & 0 & 0 & 0 & 0 & 0 & -1 \\
0 & 0 & 0 & 0 & 0 & -1 & 0
\end{array}
\right) \ .
\end{array}
\feqn
We need to look at the action of $H_o$ on $H$ to reduce the range of $x_1,\ldots, x_6$.
Starting with $\sigma$ we see that
\eqn
&& g= H(x_1,x_2,x_3,x_4,x_5,x_6) \sigma e^V \sigma H(x_9,x_{10},x_{11},x_{12},x_{13},x_{14}) \cr
&& \ = H(x_1,x_2,x_3+\frac \pi2,x_4,x_5,x_6+\frac \pi2) e^V
H(x_9 +\frac \pi2,x_{10},x_{11},x_{12} +\frac \pi2,x_{13},x_{14}).
\feqn
This shows that we can restrict $0\leq a_6 < \frac \pi2$ to avoid redundancies. A similar computation can be done for the action
of $\eta$, showing that redundancies are avoided by restricting 
$a_2\in [0,\pi/4]$. The details can be found in \cite{CCdVOS}.
So, at this time we have partially determined the ranges:
\begin{eqnarray}
&& x_1 \in [0,2\pi], \qquad x_{2} \in [0,\pi/4], \qquad x_{3}\in [0,\pi], \cr
&& x_{4} \in [0,\pi], \qquad x_{5} \in [0,\pi/2], \qquad x_{6}\in [0,\pi/2], \cr
&& x_9 \in [0,2\pi], \qquad x_{10} \in [0,\pi/2], \qquad x_{11}\in [0,\pi], \cr
&& x_{12} \in [0,\pi], \qquad x_{13} \in [0,\pi/2], \qquad x_{14}\in [0,\pi].
\end{eqnarray}
To apply the topological method we must now determine the form of the invariant measure.
This is easily computed using (\ref{measure1}), (\ref{so4param}) (and eventually the help of Mathematica):
\eqn
d\mu=27 \sqrt 3 f(2 x_7 \ , 2x_8 ) \sin (2x_2 )
\sin (2x_5) \sin (2x_{10}) \sin (2x_{13} )
\prod_{i=1}^{14} dx_i \ , \label{invmeasure}
\feqn
where
\eqn
f(\alpha, \beta ) &=&\sin (\frac {\beta-\alpha}2 ) \sin (\frac {\beta+\alpha}2 )
\sin (\frac {\beta-3\alpha}2 ) \sin (\frac {\beta+3\alpha}2 ) \sin (\alpha) \sin (\beta) \cr
&=&\frac{1}{4} (\cos(\alpha)-\cos(\beta))(\cos(3\alpha)-\cos(\beta))\sin (\alpha) \sin (\beta).
\label{ffunct}
\feqn
We see that for certain values of the angles $x_2, x_5, x_7, x_8, x_{10}, x_{13}$  the measure (\ref{invmeasure}) vanishes.
Apart from $x_7, x_8$, however, this happens only on the boundary of the chosen ranges. This means that the condition of non vanishing measure
determines the range for $x_7, x_8$ through the equation
$$
f(2 x_7 \ , 2x_8 )>0.
$$
Note that the period of $e^{x C_5}$, like the one for
$e^{\sqrt x C_{11}}$, is $2\pi$, so that we have to solve this equation
inside the square $[0,2\pi] \times [0,2\pi]$. This provides a tiling of 
the square, but it is easy to see that all the regions of such a tiling
are equivalent and we can pick any of them, see \cite{CCdVOS}. We fix:
\begin{equation}
a_7 \in [0,\pi/6] , \qquad  3a_7 \leq a_8 \leq \pi/2.
\end{equation}
Our choice for the range $R$ determines a covering $G$ of $G_2$, the volume
of which is easily computed to be:
\begin{eqnarray}
Vol(G)=\int_R d\mu =9\sqrt 3 \frac {\pi^8}{20}.
\end{eqnarray}
The final step consists in comparing this result with the expression 
obtained for the volume of $G_2$ by means of Macdonald's formula 
(\ref{macdonald}).
Indeed, the two values coincide and it holds that $Vol(G_2)=Vol(G)$. 
We can therefore infer that with our choice for
the range of the parameters, the group is covered exactly once. Instead 
of showing the details of this calculation here, we use the
$SU(3)$-parametrization determined previously to compute the volume of 
$G_2$ in yet another way. In that case, the measure was
\eqn
d\mu^{(SU(3))}_{G_2} = \frac {27}{32} \sin^5 x_6 \cos x_5 \sin^3 x_5 \sin (2x_2) d\mu_{SU(3)} dx_6 d x_5 dx_4 d x_3
dx_2 d x_1,
\feqn
so that
$$
Vol(G_2)=9\sqrt 3 \frac {\pi^8}{20} \ ,
$$
as expected.


\section{GENERALIZED EULER ANGLES FOR \boldmath{$F_4$}}
A simple construction for the Lie algebras of the exceptional Lie groups 
$F_4$ and $E_6$ is suggested by a theorem of Chevalley and Schafer \cite{CS}
which states
\begin{theorem}
The exceptional simple Lie algebra $\mathfrak{f}_4$ of dimension $52$ and rank $4$ over $K$ is the derivation algebra
$\mathfrak {D}$ of the exceptional Jordan algebra $\mathfrak {J}$ of dimension $27$ over $K$. The exceptional simple
Lie algebra $\mathfrak{e}_6$ of dimension $78$ and rank $6$ over $K$ is the Lie algebra
\eqn
\mathfrak{D}+\{R_Y\}\ , \qquad \ Tr{Y}=0 \ ,
\feqn
spanned by the derivations of $\mathfrak{J}$ and the right multiplications of elements $Y$ of trace 0.
\end{theorem}
Se also \cite{A}.
To make it workable at a practical level, we have to explain here the main 
ingredients. For our purposes, $K=\mathbb{R}$. The exceptional Jordan 
algebra is the $27$ dimensional real vector space spanned by the $3\times 3$ octonionic hermitian matrices endowed with the Abelian
product
\eq
A\circ B :=\frac 12 (A B+B A) \ ,
\feq
that is the symmetrization of the usual matrix product.\\
The derivation algebra of $\mathfrak {J}$ provides a $27$ dimensional
representation of the Lie algebra for $F_4$. However, it admits a 
decomposition in irreducible subspaces $\mathbb{R}^{27}=\mathbb{R}^{26}\oplus\mathbb{R}$, which is defined by the homomorphism:
\eqn
\ell : \mathfrak{J} \longrightarrow \mathbb{R}\ , \quad A \mapsto \sum_{i=1}^3 A_{ii} \ .\label{ell}
\feqn
Its kernel is a $26$ dimensional invariant subspace. We could restrict 
ourselves to this space, but, because the $27$ dimensional representation
can be extended to an irreducible representation of an $E_6$ algebra, we 
prefer to work with the whole space.\\
In order to concretely construct the representation, let us first realize 
an explicit isomorphism between the space of exceptional Jordan matrices
and $\mathbb{R}^{27}$:
\begin{eqnarray}
&& \Phi : \mathfrak{J} \longrightarrow \mathbb{R}^{27} \ , \quad
\left(
\begin{array}{ccc}
a_1 & o_1 & o_2 \\
o_1^* & a_2 & o_3 \\
o_2^* & o_3^* & a_3
\end{array}
\right) \longrightarrow
\left(
\begin{array}{c}
a_1 \\
\rho (o_1) \\
\rho (o_2) \\
a_2 \\
\rho (o_3) \\
a_3
\end{array}
\right) ,
\end{eqnarray}
where $a_i$, $i=1,2,3$ are real numbers, $o_i$, $i=1,2,3$ are octonions and
\begin{eqnarray}
&& \rho: \mathbb{O} \longrightarrow \mathbb{R}^8 \ , \qquad
\sum_{i=0}^7 o^i e_i \mapsto \left(
\begin{array}{c}
o^0 \\ o^1 \\ o^2 \\ o^3 \\ o^4 \\ o^5 \\ o^6 \\ o^7
\end{array}
\right).
\end{eqnarray}
In this way, the set of derivations  $\mathfrak {D}$ is mapped into the set of endomorphisms of $\mathbb{R}^{27}$.
Indeed, choosing $A_i=\Phi^{-1} (r_i)$, $r_i\in \mathbb{R}^{27}$, the identity
\begin{eqnarray}
J (A\circ B )= J (A)\circ B +A\circ J ( B )
\end{eqnarray}
provides a set of equations for the $27\times 27$ matrix $M:=\Phi J \Phi^{-1}$. This linear system can be easily solved by means of a computer, yielding
a set of $52$ linearly independent matrices. Their explicit expressions,
together with the Mathematica code generating them and their structure 
constants, can be found in \cite{BCCS1}.
We have chosen to normalize them with the conditions:
$-\frac 16 Trace (M_I M_J)=\delta_{IJ}$, $I,J=1,\ldots,52$, and $[M_i, M_j]=-\sum_{k=1}^3\epsilon_{ijk} M_k$ for $i,j\in\{1,2,3\}$.
Now, we need to recognize the $26\oplus 1$ irreducible representation. We said that this is determined by the kernel of the
map $\ell$ defined in (\ref{ell}). Composing it with the map $\Phi$,
we see that ${\rm ker} \ell \circ \Phi^{-1}=\mathbb{R} f_{27}$, where
\eqn
f_{27}=(e_1+ e_{18}+e_{27})/\sqrt {3},
\feqn
and $e_a$, $a=1,\ldots,27$ is the standard basis of $\mathbb{C}^{27}$.
Indeed, $f_{27}\in {\rm ker} M_I$ for all $I=1,\ldots,52$.\\
With respect to the new basis $\{f_a \}_{a=1}^{27}$ for $\mathbb{C}^{27}$
\eqn
&& f_1= (e_1-e_{18})/\sqrt 2 \ , \\
&& f_{18}= (e_1+e_{18}-2e_{27})/\sqrt 6 \ , \\
&& f_{27}= (e_1+ e_{18}+e_{27})/\sqrt {3} \ , \\
&& f_a= e_a\ , \mbox{  in the other cases, }
\feqn
all the matrices will have vanishing last row and column, thus explicitly
evidenciating the decomposition. We are going to call the resulting 
$27 \times 27$ matrices $\{c_i\}$, $i=1,\ldots,52$, .
The $26$ dimensional representation can then be obtained by simply 
deleting from each matrix the last row and the last column.
However, as we have remarked previously, the $27\times 27$ matrices
constitute the first $52$ elements of the $27$ dimensional fundamental
irreducible representation of $E_6$.\\
Before starting with the construction of the group, let us stop momentarily to look at some properties of the algebra.
Observe that with our matrices we can easily construct the $52$ 
dimensional adjoint representation as well. Let us call $\{C_i\}$ the 
corresponding matrices. We can easily check that the associated Killing form is negative definite and, indeed, $K_{ij}\propto \delta_{ij}$, so that we
can choose the constant to fix the Euclidean metric as the invariant metric.A possible choice for a Cartan subalgebra is
$H=\mathbb{R}C_1\oplus\mathbb{R} C_6 \oplus\mathbb{R} C_{15} \oplus\mathbb{R} C_{36}$ and the roots can then be computed by simultaneously
diagonalizing the generators $\{C_a\}$, $a=1, 6, 15, 36$:
$$
C_a \vec v_i =\lambda_{a,i} \vec v_i,\qquad i=1,\ldots, 27, \quad\ \vec v_i\in \mathbb{C}^{52}.
$$
The resulting vectors $(\lambda_{1,i}, \lambda_{6,i}, \lambda_{15,i}, \lambda_{36,i})$, $i=1,\ldots, 27$, represent the roots, which, indeed,
coincide with the roots of $F_4$, as expected, thus proving that we have 
obtained a realization of the compact form of the $F_4$ Lie algebra.\\
To construct the corresponding group, it is useful to identify its 
subalgebras first. By studying the commutators, we see that the first $21$ 
matrices generate a $\mathfrak{so}(7)$ subalgebra, whose 
$\mathfrak{so}(i)$ subalgebras, with $i=6,5,4,3$, are generated by the 
first $i(i-1)/2$ matrices, respectively.
A possible choice for the relative Cartan subalgebras is $C_1$ for 
$\mathfrak{so}(3)$; $C_1,C_6$ for $\mathfrak{so}(4)$
and $\mathfrak{so}(5)$ and $C_1,C_6,C_{15}$ for $\mathfrak{so}(6)$ and $\mathfrak{so}(7)$. This can be used to compute the corresponding roots and to 
check the algebras. Adding to $\mathfrak{so}(7)$ the matrices $c_i$, 
with $i=30,\ldots,36$, we obtain a $\mathfrak{so}(8)$ subalgebra. This is
the Lie algebra associated to the $Spin(8)$ subgroup of $F_4$ which leaves 
invariant the three Jordan matrices $J_i$, $i=1,2,3$, where $J_i$ is the
matrix which has $\{J_i\}_{ii}=1$ as the unique non-vanishing entry. 
Indeed, we find that for $i=1,2,3$, $\Phi(J_i)$ belongs to the kernel of 
the $\mathfrak{so}(8)$ matrices. \\
The $\mathfrak{so}(8)$ algebra can be extended to a $\mathfrak{so}(9)$ subalgebra in three different ways:
first, the algebra $\mathfrak{so}(9)_{_1}$ obtained by adding 
$c_{45},\ldots,c_{52}$ to $\mathfrak{so}(8)$, and corresponding to the 
subgroup $Spin(9)_{_1}$ of $F_4$ which leaves $J_1$ invariant; second,
the algebra $\mathfrak{so}(9)_{_2}$ obtained by adding
$c_{37},\ldots,c_{44}$ to $\mathfrak{so}(8)$, and corresponding to the 
subgroup $Spin(9)_{_2}$ of $F_4$ which leaves $J_2$ invariant; finally
the $\mathfrak{so}(9)_{_3}$ obtained by adding $c_{22},\ldots,c_{29}$ to 
$\mathfrak{so}(8)$, and corresponding to the subgroup $Spin(9)_{_3}$ 
of $F_4$ which leaves $J_3$ invariant.
Again, this can be checked by applying the given matrices to $\Phi (J_1)$, $\Phi(J_2)$ and $\Phi(J_3)$ respectively. We will use $Spin(9)_{_1}$ which we will refer to simply as $Spin(9)$.\\
Finally, recall that if $\mathfrak {p}$ is the linear complement of $\mathfrak{so}(9)$ in $F_4$, from $ad$-invariance
and orthogonality it follows that:
\eqn
&& [\mathfrak{so}(9),\mathfrak{p}] \subset \mathfrak{p} \ , \\
&& [\mathfrak{p},\mathfrak{p}]\subset \mathfrak{so}(9) \ .
\feqn

\subsection{The generalized \boldmath{$Spin(9)$}-Euler construction.}


\subsubsection{The maximal subgroup}
In this section we start with the construction of the Euler parametrization
for $F_4$, based on its maximal subgroup $H=Spin(9)$. In particular, 
out of the three $Spin(9)$ subgroups we have been able to identify 
previously, we pick $Spin(9)_{_1}$.
Then, its complementary subalgebra $\cP$ is the $16$ dimensional real 
vector space generated by the matrices $c_i$, with 
$i=22,\ldots,29,37,\ldots,44$. A look at the structure constants shows 
that as subspace $V$ (see (\ref{eulero})) we can take any 1-dimensional 
subspace of $\cP$.
We choose $c_{22}$ as the generator for $V$. Since ${\rm dim} G- {\rm dim} H -{\rm dim} V=21$ we expect for $H_o$ to be a
$Spin(7)$ subgroup of $Spin(9)$.
To check that this is true, let us first recall that first $21$ matrices
generate an $so(7)$ algebra. We are now able to construct a new set of 
$21$ generators $\{\tilde c_i\}$, $i=1,\ldots,21$, which commute with
$c_{22}$ and which have the same structure constants as the $\{c_i\}$. 
To this end we start with the $so(8)$ subalgebra
generated by $\{c_I\}$, $I=1,\ldots,21,30,\ldots,36$. Then the matrices
$\{c_\alpha\}$, $\alpha=30,\ldots,36$, generate the whole $so(7)$ algebra
through:
\eqn
c_{\frac {k(k-1)}2+i+1}=[c_{30+i},c_{30+k}] \ , k=1,\ldots,6, \ i=0,\ldots,k-1 \ .
\feqn
Notice that for $a,b\in\{22,\ldots,29\}$ the commutator $[c_a,c_b]$ is a 
combination of four elements of $so(8)$, all having the
same commutator with $c_{22}$. With this in mind, let us define
\eqn
\tilde c_{30+i}:=-[c_{22},c_{23+i}] \ , \quad i=0,\ldots,7 \ ,
\feqn
and then
\eqn
\tilde c_{\frac {k(k-1)}2+i+1}=[\tilde c_{30+i},\tilde c_{30+k}] \ , k=1,\ldots,6, \ i=0,\ldots,k-1 \ .
\feqn
Thus, the matrices $\{\tilde c_I \}$, $I=1,\ldots,21$, $30,\ldots,36$ 
have exactly the same structure constants as the $\{c_I\}$ and 
$[\tilde c_i, c_{22}]=0$ for $i=1,\ldots,21$. This is the 
$\mathfrak{so}(7)$ we were looking for, let us call it
$\mathfrak {h}_o$, so that $H_o=\exp (\mathfrak {h}_o)$. 
In order to apply (\ref{eulero}) we need to have an explicit expression
for $H$ first. This can be done by applying to it the same method we are 
using for $F_4$, i.e. by constructing its Euler parametrization with
$SO(8)$ as a maximal subgroup, which in turn can be constructed from its 
$SO(7)$ maximal subgroup and so on, with an inductive procedure.
To avoid annoying repetitions to the reader, we limit ourselves here to
the final expression for $H$:
\eqn
&& Spin(9)[x_1,\ldots,x_{36}]= e^{x_1 c_3} e^{x_2 c_{16}} e^{x_3 c_{15}} e^{x_4 c_{35}} e^{x_5 c_5} e^{x_6 c_1} e^{x_7 c_{30}} e^{x_8 c_{45}}
e^{x_9 c_3} e^{x_{10} c_{16}} e^{x_{11} c_{15}}\cr
&& {\phantom {Spin(9)[x_1,\ldots,x_{36}]=}}
e^{x_{12} c_{35}} e^{x_{13} c_5} e^{x_{14} c_1} e^{x_{15} c_{30}}  e^{x_{16} c_3} e^{x_{17} c_5} e^{x_{18} c_4} e^{x_{19} c_7} e^{x_{20} c_{11}}
 e^{x_{21} c_{16}}
\cr && {\phantom {Spin(9)[x_1,\ldots,x_{36}]=}} e^{x_{22} c_3} e^{x_{23} c_5} e^{x_{24} c_4} e^{x_{25} c_7} e^{x_{26} c_{11}}
e^{x_{27} c_3} e^{x_{28} c_5} e^{x_{29} c_4} e^{x_{30} c_7} e^{x_{31} c_3} e^{x_{32} c_5}  \cr
&& {\phantom {Spin(9)[x_1,\ldots,x_{36}]=}} e^{x_{33} c_4} e^{x_{34} c_3} e^{x_{35} c_2} e^{x_{36} c_3}, \label{spin9}
\feqn
with ranges
\eqn
&& x_i \in [0,2\pi], \qquad i=1,2,3,9,10,11,16,22,27,31,34, \cr
&& x_i \in [0,\pi], \qquad i=4,8,12,17,21,23,26,28,30,32,33,35,  \cr
&& x_i \in [-\frac \pi2,\frac \pi2], \qquad i=5,13,18,19,20,24,25,29, \cr
&& x_i \in [0,\frac \pi2], \qquad i=6,7,14,15, \cr
&& x_{36} \in [0,4\pi],
\feqn
and measure
\eqn
&& d\mu_{_{Spin(9)}}[x_1,\ldots,x_{36}] =\sin x_4 \cos x_5 \cos x_6 \sin^2 x_6 \cos^4 x_7 \sin^2 x_7 \sin^7 x_8 \cr
&& {\phantom {d\mu_{_{Spin(9)}}[x_1,\ldots,x_{36}] =}} \sin x_{12} \cos x_{13} \cos x_{14} \sin^2 x_{14} \cos^2 x_{15} \sin^4 x_{15}\cr
&& {\phantom {d\mu_{_{Spin(9)}}[x_1,\ldots,x_{36}] =}} \sin x_{17} \cos^2 x_{18} \cos^3 x_{19} \cos^4 x_{20} \sin^5 x_{21} \cr
&& {\phantom {d\mu_{_{Spin(9)}}[x_1,\ldots,x_{36}] =}} \sin x_{23} \cos^2 x_{24} \cos^3 x_{25} \sin^4 x_{26} \cr
&& {\phantom {d\mu_{_{Spin(9)}}[x_1,\ldots,x_{36}] =}} \sin x_{28} \cos^2 x_{29} \sin^3 x_{30} \sin x_{32} \sin^2 x_{33} \sin x_{35}
\prod_{i=1}^{36} dx_i.
\feqn

\subsubsection{The whole \boldmath{$F_4$}}
To construct the quotient $B$ we need to identify the subgroup $SO(7)$ in $H$. We have seen that this group is generated by the matrices
$\{\tilde c_i\}$, $i=1,\ldots,21$. We have also seen that these matrices
satisfy the same commutation relation as the matrices $\{c_i\}$. Thus, 
if we are able to extend this to the whole $\mathfrak{so}(9)$ algebra, we
can use the same expression (\ref{spin9}) with the matrices
$\{\tilde c_i\}$ instead. Luckily, we see that it is enough to add the 
matrices
\eqn
\tilde c_i=c_{i+8}, \qquad\ i=22,\ldots,28,\cr
\tilde c_i=c_{i+16}, \qquad\ i=29,\ldots,36,
\feqn
to obtain the desired set $\{\tilde c_a\}$, $a=1,\ldots,36$ generating 
the whole $Spin (9)$ group. Because the last $21$ exponentials 
in (\ref{spin9}) generate the $SO(7)=H_o$ group, we get
\eqn\label{B}
&& B[x_1,\ldots, x_{15}]=e^{x_1 \tilde c_3} e^{x_2 \tilde c_{16}} e^{x_3 \tilde c_{15}} e^{x_4 \tilde c_{35}} e^{x_5 \tilde c_5} e^{x_6 \tilde c_1}
e^{x_7 \tilde c_{30}} e^{x_8 \tilde c_{45}} e^{x_9 \tilde c_3} e^{x_{10} \tilde c_{16}} e^{x_{11} \tilde c_{15}} \cr
&& {\phantom{B[x_1,\ldots, x_{15}]=}} e^{x_{12} \tilde c_{35}} e^{x_{13} \tilde c_5} e^{x_{14} \tilde c_1} e^{x_{15} \tilde c_{30}} .
\feqn
Therefore, the resulting Euler parametrization of $F_4$ is:
\eqn
F_4[x_1,\ldots,x_{52}]=B[x_1,\ldots,x_{15}]e^{x_{16} c_{22}} Spin(9)[x_{17},\ldots,x_{52}] \ .
\feqn
Here, the range for the parameters $x_1,\ldots,x_{16}$ remains to be
determined, while the other ranges are the ones for $Spin(9)$.
We need to apply the topological method.
To this purpose, we need to compute ${\rm det} (J_p)$ as in 
(\ref{measure1}). This computation is quite involved, and it requires some 
technical trick to be performed. We refer to \cite{BCCS1} for the details.
The resulting measure is:
\eqn
&& d\mu_{F_4}[x_1,\ldots,x_{52}]=d\mu_o [x_1,\ldots,x_{16}] d\mu_{Spin(9)}[x_{17},\ldots,x_{52}],\\
&& d\mu_o [x_1,\ldots,x_{16}]=2^7 \cos^7 \frac {x_{16}}2  \sin^{15} \frac {x_{16}}2 \sin x_4 \cos x_5 \cos x_6 \sin^2 x_6 \cos^4 x_7 \sin^2 x_7 \sin^7 x_8 \cdot \cr
&& \qquad\ \cdot\sin x_{12} \cos x_{13} \cos x_{14} \sin^2 x_{14} \cos^2 x_{15} \sin^4 x_{15} \prod_{i=1}^{16} dx_i \ .
\feqn
{From} this we can select the ranges along the lines explained in section \ref{sec:topo}.
Note that the exponentials are trigonometric functions of $x_i/2$ with 
periods $4\pi$, so that we should take the range $x_i=[0,4\pi]$ for 
$i=1,2,3$ and $i=9,10,11$.
However, for all the $\tilde c_i \in so(7)$ we have that 
$e^{2\pi \tilde c_i}$ commute with $\tilde c_j$ and
with $c_{22}$, so that it can be reabsorbed in the $Spin(9)$ factor of $F_4$ and these periods can be reduced to $[0,2\pi]$.
The ranges determined by the topological method are then
\eqn
&& x_1 \in [0,2\pi] \ , \quad x_2 \in [0,2\pi] \ , \quad x_3 \in [0,2\pi] \ , \quad x_4 \in [0,\pi] \ , \cr
&& x_5 \in [-\frac \pi2,\frac \pi2] \ , \quad x_6\in [0,\frac \pi2] \ , \quad x_7\in [0,\frac \pi2] \ , \quad x_8\in [0,\pi] \ , \cr
&& x_9 \in [0,2\pi] \ , \quad x_{10} \in [0,2\pi] \ , \quad x_{11} \in [0,2\pi] \ , \quad x_{12} \in [0,\pi] \ , \cr
&& x_{13} \in [-\frac \pi2,\frac \pi2] \ , \quad x_{14}\in [0,\frac \pi2] \ , \quad x_{15}\in [0,\frac \pi2] \ ,
\quad x_{16}\in [0,\pi] \ .
\feqn
This choice of the range covers the whole group at least once. Let us call 
$M$ the corresponding homological cycle.
Integrating the measure on the full range we obtain
\eqn
\mu(M)=\frac {2^{26} \cdot \pi^{28}}{3^7 \cdot 5^4 \cdot 7^2 \cdot 11}.
\feqn
To be sure that we covered $F_4$ exactly once, we must compute the volume of the group by means of the Macdonald formula.
Its Betty numbers were computed in \cite{chevalley}. For $F_4$ there are four free generators for the rational homology, corresponding to
four spheres having dimensions
\eqn
&& d_1=3, \ d_2=11, \ d_3=15, \ d_4=23 \ .
\feqn
They contribute to the volume with a term
\eqn
Vol(S^3) Vol(S^{11}) Vol(S^{15}) Vol(S^{21}) =2{\pi^2} 2\frac {\pi^5}{4!} 2 \frac {\pi^7}{6!} 2 \frac {\pi^{11}}{10!} .
\feqn
The simple roots are \cite{FH}
\eqn
&& r_1 =L_2-L_3 \\
&& r_2 =L_3-L_4 \\
&& r_3 =L_4 \\
&& r_4 =\frac {L_1-L_2-L_3-L_4}2
\feqn
where $L_i$, $i=1,\ldots,4$ is an orthonormal base for the dual Cartan algebra.
The volume of the fundamental region representing the torus is then $1/2$.
Finally, there are $48$ non vanishing roots, $24$ of with length $1$, 
and $24$ with length $\sqrt 2 $. We have determined them explicitly 
and, as expected, they correspond to the ones just presented, with 
$L_i=e_i$, the canonical basis of $\mathbb{R}^4$.
The resulting contribution is the term:
\eqn
\prod_{\alpha \in R(F_4)} \frac 2{|\alpha|}= 2^{48}{(\sqrt 2)^{24}}.
\feqn
The volume of the group is then
\eqn
Vol(F_4) =\frac {2^{26} \cdot \pi^{28}}{3^7 \cdot 5^4 \cdot 7^2 \cdot 11} \ . \label{volumeF4}
\feqn
We conclude that the range we have determined covers the group 
exactly a single time.

\section{THE \boldmath{$F_4$}-EULER ANGLES FOR \boldmath{$E_6$}}
As the construction of the Euler parametrization for $E_6$ is very 
similar to the one for $F_4$, we are going to be very short and refer to
\cite{BCCS2} for more details.
We can use the theorem of Chevalley and Schafer previously cited to 
extend the representation of the $F_4$ algebra to the 27 irreducible 
representation of the whole $E_6$ algebra, by simply adding the 
the matrices representing the action of $R_Y$.
We only need to associate a $27\times 27$ matrix $M(A)$ to each $A\in \mathfrak{J}$, in such the way that, if $v\in \mathbb{R}^{27}$, then
\begin{equation}
M(A) v=\Phi (A\circ \Phi^{-1} (v)).
\end{equation}
The set of traceless Jordan matrices being $26$-dimensional, this adds 
$26$ new generators, which complete the $F_4$ algebra to the
$78$-dimensional $E_6$-algebra. However, by computing the Killing form we
can easily check that this is not the compact form with signature
$(52,26)$. It is instead the non compact form $E_{6(-26)}$. 
Fortunately, we can obtain the compact form by multiplying the $26$ 
generators we have added by $i$. In this way, the algebra remains real 
and the representation becomes complex, and now $V=\mathbb{C}^{27}$.
We haverealized these matrices with Mathematica and in the basis
\begin{eqnarray}\label{jordan}
\left(
\begin{array}{ccc}
a_1 & o_1 & o_2 \\
o_1^* & a_2 & o_3 \\
o_2^* & o_3^* & -a_1 -a_2
\end{array}
\right)
\end{eqnarray}
for the traceless Jordan matrices.
They can be found in \cite{BCCS2}.\\
As the next step, we now need to choose a maximal compact subgroup.
It is convenient to select the largest one, which we know to be 
$H=F_4$, in our case the group generated by the firsts $52$ matrices.
Its linear complement $\cP$ (in the algebra) contains two preferred 
elements associated to the two diagonal matrices (\ref{jordan}) 
with $a_1=1, a_2=0$ and $a_1=0, a_2=1$, respectively. Following
the order dictated by the map $\Phi$, after orthonormalization w.r.t. the product $(J|J')={\rm Trace} (J\circ J')$ in $\mathfrak{J}$,
these will correspond to the matrices $c_{53}$ and $c_{70}$ respectively. This is indeed the expression
we used in \cite{BCCS2} to do the computer calculations. There, we have found convenient a posteriori to recombine these two matrices
in the new generators
\begin{eqnarray*}
{\tilde c}_{53}=\frac 12 c_{53} +\frac {\sqrt 3}2 c_{70},\\
{\tilde c}_{70}=-\frac {\sqrt 3}2 c_{53} +\frac 12 c_{70}.
\end{eqnarray*}
These, added to the four matrices previously considered for the Cartan 
subalgebra of $F_4$, generate a Cartan subalgebra of
$E_6$ and the corresponding roots are exactly the ones described, for 
example, in \cite{FH}, with $L_i$ replaced by the elements $e_i$
of the standard basis of $\mathbb {R}^6$.\\
In any case, it is easy to check that ${\tilde c}_{53}, {\tilde c}_{70}$ can be taken as generators of $V$. Obviously, they commute.
To realize the Euler parametrization, we note that the redundancy is now $28$-dimensional, so that we expect to find a $28$ dimensional
subgroup $H_o$ of $H$ which commutes with $V$. In fact, this happens to be
the $SO(8)$ subgroup generated by the first $28$ matrices
$\{c_i\}$, $i=1,2,\ldots,28$.
We can then write
$$
E_6[x_1,\ldots,x_{78}]=B_{E_6}[x_1,\ldots,x_{24}] e^{x_{25} c_{53}+x_{26} c_{70}} F_4 [x_{27},\ldots,x_{78}] \ ,
$$
with $B_{E_6}=F_4 / SO(28)$ and $F_4$ as in the previous section. This means that in particular
\eqn
B_{E_6}[x_1,\ldots,x_{24}]=B[x_1,\ldots,x_{15}]e^{x_{16} c_{22}} B_9[x_{17},\ldots,x_{23}]e^{x_{24}c_{37}} \ ,
\feqn
where $B$ is given by (\ref{B}) and
\begin{eqnarray*}
B_9[x_1,\ldots,x_7] =e^{x_1 \tilde c_3} e^{x_2 \tilde c_{16}} e^{x_3 \tilde c_{15}} e^{x_4 \tilde c_{35}} e^{x_5 \tilde c_5}
e^{x_6 \tilde c_1} e^{x_7 \tilde c_{30}}.
\end{eqnarray*}
We can now compute the associated invariant measure. The calculation is quite involved and details can be found in \cite{BCCS2}. Here we
give the final result only:
\eqn
&& d\mu_{E_6}=2^7 \sin x_4 \cos x_5 \cos x_6 \sin^2 x_6 \cos^4 x_7 \sin^2 x_7 \sin^7 x_8 \cdot \cr
&& \qquad\ \ \sin x_{12} \cos x_{13} \cos x_{14} \sin^2 x_{14} \cos^2 x_{15} \sin^4 x_{15}\ \cos^{15} \frac {x_{16}}2
\sin^7 \frac {x_{16}}2  \cdot \cr
&& \qquad\ \ \sin x_{20} \cos x_{21} \cos x_{22} \sin^2 x_{22} \cos^2 x_{23} \sin^4 x_{23} \sin^7 x_{24} \cdot \cr
&& \qquad\ \ \sin^8 x_{25}
\sin^8 \left( \frac {\sqrt 3}2 x_{26} +\frac {x_{25}}2 \right) \sin^8 \left( \frac {\sqrt 3}2 x_{26} -\frac {x_{25}}2 \right)\cdot \cr
&& \qquad\ \ d\mu_{F_4}[x_{27},\ldots,x_{78}] \prod_{i=1}^{26} dx_i.
\feqn
Proceeding as for $F_4$, from this measure we can determine the range $R$ for the parameters:
\eqn
&& x_1 \in [0,2\pi] \ , \quad x_2 \in [0,2\pi] \ , \quad x_3 \in [0,2\pi] \ , \quad x_4 \in [0,\pi] \ , \cr
&& x_5 \in [-\frac \pi2,\frac \pi2] \ , \quad x_6\in [0,\frac \pi2] \ , \quad x_7\in [0,\frac \pi2] \ , \quad x_8\in [0,\pi] \ , \cr
&& x_9 \in [0,2\pi] \ , \quad x_{10} \in [0,2\pi] \ , \quad x_{11} \in [0,2\pi] \ , \quad x_{12} \in [0,\pi] \ , \cr
&& x_{13} \in [-\frac \pi2,\frac \pi2] \ , \quad x_{14}\in [0,\frac \pi2] \ , \quad x_{15}\in [0,\frac \pi2] \ ,
\quad x_{16}\in [0,\pi] \ , \cr
&& x_{17} \in [0,2\pi] \ , \quad x_{18} \in [0,2\pi] \ , \quad x_{19} \in [0,2\pi] \ , \quad x_{20} \in [0,\pi] \ , \cr
&& x_{21} \in [-\frac \pi2,\frac \pi2] \ , \quad x_{22}\in [0,\frac \pi2] \ , \quad x_{23}\in [0,\frac \pi2] \ , \quad x_{24}\in [0,\pi] \ , \cr
&& x_{25}\in [0,\frac \pi2] \ , \quad -\frac {x_{25}}{\sqrt 3} \leq x_{26} \leq \frac {x_{25}}{\sqrt 3} \ ,
\feqn
and $x_j$, $j=27,\ldots,78$, chosen to cover the whole $F_4$ group. This choice defines a $78$ dimensional closed cycle $W$ having volume
\eqn
Vol (W)=\int_R d\mu_{E_6} =\frac {\sqrt 3 \cdot 2^{17} \cdot \pi^{42}}{3^{10} \cdot 5^5 \cdot 7^3 \cdot 11}.
\feqn
To complete the work we need to check that this is indeed the volume of $E_6$ as given by the Macdonald formula.
The rational homology of $E_6$ is $H_*(E_6)=H_*(\prod_{i=1}^6 S^{d_i})$, with (\cite{chevalley})
\eqn
d_1=3, \ d_2=9, \ d_3=11, \ d_4=15, \ d_5=17, \ d_6=23\ .
\feqn
$E_6$ is simply laced, with simple roots
\eqn
&& r_1 =L_1+L_2 \\
&& r_2 =L_2-L_1 \\
&& r_3 =L_3-L_2 \\
&& r_4 =L_4-L_3 \\
&& r_5 =L_5-L_4 \\
&& r_6 =\frac {L_1-L_2-L_3-L_4-L_5+\sqrt 3 L_6}2
\feqn
where $L_i$, $i=1,\ldots,6$ is an orthogonal basis for the dual of the Cartan algebra. The volume of the torus associated to it is then $\frac 2L$. 
As a check for the algebra, we have explicitly verified that the 
$72$ roots of the algebra coincide with the roots of $E_6$,
each one having length $\sqrt 2$. They have, indeed, the structure 
given in \cite{FH}, with $L_i=e_i$, the canonical basis of
$\mathbb{R}^6$. The Macdonald formula then provides the result
\eqn
Vol(E_6) =\frac {\sqrt 3 \cdot 2^{17} \cdot \pi^{42}}{3^{10} \cdot 5^5 \cdot 7^3 \cdot 11},
\feqn
which concludes our check.

\section{CONSTRUCTION OF NON COMPACT SPLIT FORMS AND THEIR COSET MANIFOLDS}
Up to now we have considered compact groups only. However, as discussed in the introduction, it is important to be able to concretely
realize non compact groups also. A particular class is given by the split 
forms, for which a particularly suitable technique is the Iwasawa 
decomposition, that we are going to discuss in this section.
In order to clearly show the advantage of such method, in the next section
we are going to compare the construction of a non compact form obtained
by analytic continuation of a compact one with the direct Iwasawa
construction.

\subsection{Analytic continuation of the generalized Euler angles}
A first way to realize a split by starting from the compact one is the following. Suppose we have realized a Euler parametrization of the compact 
group $G$ with respect to a maximal subgroup $H$, say
$$
G[x_1,\ldots,x_p; y_1,\ldots,y_r;z_1,\ldots,z_m]=B[x_1,\ldots,x_p] e^{V[y_1,\ldots,y_r]} H[z_1,\ldots,z_m], \qquad\ p+r+m=n.
$$
This is based on the orthogonal decomposition
$\mathfrak {g}=\mathfrak {h}+\mathfrak {p}$. We know that
$[\mathfrak {h},\mathfrak {h}]\subset \mathfrak {h}$ and $[\mathfrak {h},\mathfrak {p}]\subset \mathfrak {p}$. Let us now suppose
that the further condition $[\mathfrak {p},\mathfrak {p}]\subset \mathfrak {h}$ is satisfied. This condition is also called {\it symmetry}.
It is a non trivial condition and it requires $\mathfrak {h}$ to be 
a maximal subalgebra. Indeed, suppose we start with such a decomposition 
and we fix a subgroup $H'\subset H$.
This determines the new orthogonal decomposition
$$
\mathfrak {g}=\mathfrak {h}'+\mathfrak {p}'=\mathfrak {h}'+\mathfrak {p}+\mathfrak {p}^{\prime\prime},
$$
with $\mathfrak {p}^{\prime\prime}=\mathfrak {p}'\cap \mathfrak {h}$. Thus,
$$
[\mathfrak {p},\mathfrak {p}^{\prime\prime}] \subset \mathfrak {p}\subset \mathfrak {p}'
$$
violates symmetry.\\
On the other hand we can easily see that symmetry is satisfied by all examples we considered, and indeed this happens to be true
for all simple Lie groups \cite{G}. Therefore, we can go from the 
compact form to the non compact form corresponding to the given maximal
subgroup, simply by the {\it Weyl unitary trick} \cite{G,FH}:
$$
\mathfrak {p}\longmapsto i\mathfrak {p},
$$
$i$ being the imaginary unit. Thus, the Euler parametrization of the split form is given by
$$
G_{\rm split}[x_1,\ldots,x_p; y_1,\ldots,y_r;z_1,\ldots,z_m]=G[i x_1,\ldots,i x_p; i y_1,\ldots,i y_r;z_1,\ldots,z_m].
$$

\subsection{The Iwasawa decomposition}
While the Euler decomposition is particularly suitable for realizing 
compact Lie groups, for the non compact split forms a much simpler 
realization is provided by the Iwasawa decomposition\cite{I}. It
is based on the Cartan decomposition relative to a maximal subgroup $H$, 
$\mathfrak{g}=\mathfrak{h}\oplus \mathfrak{s}$, where
$\mathfrak{h}$ is the Lie algebra associated to $K$ and $\mathfrak{s}$
its linear complement. The Cartan decomposition requires the existence 
of a linear involution
$\theta: \mathfrak{g}\longrightarrow \mathfrak{g}$ such that restricted to
$\mathfrak{s}$ the quadratic form
$$
B:\mathfrak{s}\times \mathfrak{s} \longrightarrow \mathbb{R},\quad (a,b) \longmapsto B(a,b)=K(a,\theta(b))
$$
is positive definite.

Recall that the Killing form on the compact form is negative definite. Starting from our orthogonal decomposition
$\mathfrak{g}=\mathfrak{p}\oplus \mathfrak{h}$ we see that the map
$$
\theta: \mathfrak{g}\longrightarrow \mathfrak{g},\quad (a,b)\mapsto (a,-b), \quad \forall (a,b)\in \mathfrak{p}\oplus \mathfrak{h},
$$
satisfies the required conditions so that we can identify 
$\mathfrak{p}$ with $\mathfrak{s}$.\\
The next step consists in selecting a Cartan subalgebra of $\mathfrak{p}$.
We call it $\mathfrak{a}$ and $A$ the group it generates. 
Being a Cartan subalgebra, the adjoint action of $\mathfrak{a}$ is 
diagonalizable and we can associate to it a complete set of positive roots.
{From} the basic properties of the root spaces, we know that the 
corresponding eigenmatrices generate a nilpotent subalgebra $\mathfrak{n}$.
The Iwasawa decomposition states that the non compact form of $G$ 
associated to the maximal subgroup $H$ can be realized as
\begin{eqnarray}
G=HAN=e^{\mathfrak{h}} e^{\mathfrak{a}} e^{\mathfrak{n}}.
\end{eqnarray}
Note that since $H$ is a compact group, we can use our Euler 
parametrization to describe it.
Then all new information is contained in the non compact quotient $G/H$.
Before investigating how this can be described, let us make some further
comments on the comparison between the Euler and Iwasawa constructions.
On one hand, there doesn't exist any compact counterpart of the Iwasawa 
construction, but surely there are many other possibilities, as for 
example the exponential map itself. In this case, the big advantage of the
Euler construction is that involves only parametric angles, which appear 
in the expression for the group elements in a trigonometric form.
Computationally, this is not immediately an advantage because of the 
difficulties in handling trigonometric simplifications with Mathematica.
Indeed, at some steps direct manipulations of the expressions by hand
has been necessary and in fact much simpler than direct computer
computing. However, the true advantage arises when the explicit range of
the parameters has to be established. For this purpose, as we have seen,
the periodicity of the trigonometric expressions provides a quite direct
way to determine such ranges, whereas for a generic parametrization this
would require the solution of some transcendental equations, which can be 
handled only numerically.\\
On the other hand, when we work with a non compact form, the difficult
problem of determining the explicit range for the parameters is restricted
to the compact subgroup only. Therefore, we need to use the trigonometric
expressions for the compact subgroup only, while it is now possible to use 
a simpler realization for the non compact part, possibly much easier to
handle. Such a realization can be provided exactly by the Iwasawa
decomposition, where only Abelian or nilpotent matrices appear in the
non compact part.
In particular, from the structure of the root spaces the nilpotency of the
non compact part will be at most the rank $r$ of the group, so that we expect for the non compact part to appear polynomial terms
of degree at most $r$, instead of trigonometric expressions (or hyperbolic after the Weyl trick).

\subsection{The coset manifold}
Let  us now look at the construction of the non compact quotient $G/K$. We need to compute the induced metric (\ref{quoz}) starting from
the Iwasawa expression. Obviously, we can proceed exactly as for the
compact case. However, following the tradition, we have written the
decomposition taking $H$ as a left factor instead of a right factor, so
that it will be convenient here to exchange left invariant forms
with right invariant form. This does not change the substance, being the 
Killing form bi invariant. Now, let us introduce the one form
\begin{eqnarray*}
&& J^R_G=dG \ G^{-1}= HA dN\ N^{-1} A^{-1} H^{-1}+ HdA\ A^{-1} H^{-1} +dH\ H^{-1}\cr
&& \quad\ \equiv HA J_N A^{-1} H^{-1} +H J_A H^{-1}+J_H.
\end{eqnarray*}
To compute the metric of $M=G/H$, we need to eliminate the components of $J^R_G$ along the fibers ($\mathfrak{h}$), so as to define the reduced
form $J'_G$, giving the metric
$$
d\sigma^2=\kappa {\rm Tr} ({J'_G \otimes J'_G}),
$$
where $\kappa$ is a normalization constant.
Let us study the structure of this metric. First, notice that the term 
$J_H$, which appears in $J^R_G$, is projected out to obtain $J'_G$.
Moreover, the adjoint action of $H$ commutes with the projection, because
it respects the direct decomposition
$\mathfrak{g}=\mathfrak{h}\oplus \mathfrak{p}$. Therefore, if we define
$$
J_p:=\pi(A J_N A^{-1}),
$$
where $\pi$ is the projection out from the fibers, then
$$
d\sigma^2=\kappa {\rm Tr} ({J_p \otimes J_p})+ \kappa {\rm Tr} (J_A\otimes J_A).
$$
Let us remark that $J_p$ is orthogonal to $J_A$. Indeed,
$J_A$ is easily computed, because $A=\exp (\sum_{i=1}^r) y_i H_i$ defines 
an Abelian group. Here $H_i$ identifies an orthonormal basis 
(with respect to the product $\kappa {\rm Tr}$) for the Cartan subspace,
so that $J_A=\sum_{i=1}^r) dy_i H_i$ and
$$
\kappa {\rm Tr} (J_A\otimes J_A)=\sum_{i=1}^r) dy_i^2.
$$
On the other side, $J_p$ can be easily determined from the simple
properties of $N$. Recall that the generators of $N$ are the positive
root matrices $R_l$, $l=1,\ldots,m:=(n-r)/2$, so that two such matrices 
commute if the sum of the corresponding roots is not a root, otherwise the
commutator is proportional to the matrix associated to the resulting root.
Now, $N(x_1,\ldots,x_m)=e^{\sum_{i=1^m} x_i R_i}$, and
\begin{eqnarray}
J_N= \sum_{i=1}^m n^i(\vec x) R_i, \qquad\ n^i (\vec x)=\sum_{j=1}^m n^i_{\ j} (\vec x) dx_j \ ,
\end{eqnarray}
where the $n^i_j(x_1,\ldots, x_m)$ are all polynomials in the $x_i$.
Now, the $R_i$ are eigenmatrices for the action of $A$ and, therefore,
we have
\begin{eqnarray}
A R_i A^{-1}= e^{\sum_{a=1}^r r_{i,a} y_a} R_i,
\end{eqnarray}
where $\vec r_i=(r_{i,1}, \ldots, r_{i,r})$ are the positive roots whose components are the eigenvalues $r_{i,a}$, $a=1,\ldots,r$ of $H_a$
with eigenvector $R_i$.
Thus,
\begin{eqnarray}
A \, J_n \, A^{-1}=\sum_{i=1}^m e^{\sum_{a=1}^r r_{i,a} y_a} n^i(\vec x) R_i,
\end{eqnarray}
and to obtain $J_p$ we only need to take the projection on $\mathfrak{p}$.
The metric on the quotient is then
\begin{eqnarray}
&& d\sigma^2= \sum_{i=1}^r dy_i^2 +\sum_{i=1}^r e^i\otimes e^i,\\
&& e^i=\kappa {\rm Tr} [A \, J_n \, A^{-1} P_i],
\end{eqnarray}
where $P_i$, together with $H_a$ realize an orthonormal basis of $\mathfrak {p}$ with respect to the product $(a|b)=\kappa {\rm Tr}(ab)$.

\section{REALIZING \boldmath{$G_{2(2)}$} AND \boldmath{$G_{2(2)}/SO(4)$}}
The non compact form $G_{2(2)}$ is the split form of $G_2$ associated to the maximal compact subgroup $SO(4)$. Referring to section \ref{sec:g2},
we know that $SO(4)$ is generated by $C_i$, $i=1,2,3,8,9,10$, and $\mathfrak {p}$ is generated by $C_a$, $a=4,5,6,7,11,12,13,14$.
To determine the split form we could multiply $C_a$ by the imaginary unit $i$. Alternatively, noting that all matrices are antisymmetric,
we prefer to transform the matrices $C_a$ into symmetric matrices.
The representative matrices obtained in this way
are normalized with the condition $Tr (Q_I Q_J ) = \eta_{IJ}$, where $\eta={\rm diag}\{-1,-1,-1,1,1,1,1,-1,-1,-1,1,1,1,1\}$:
$$
Q_1 =\left(
\begin{array}{ccccccc}
0 & 0 & 0 & 0 & 0 & 0 & 0 \\
0 & 0 & 0 & 0 & 0 & 0 & 0 \\
0 & 0 & 0 & 0 & 0 & 0 & 0 \\
0 & 0 & 0 & 0 & 0 & 0 & -1 \\
0 & 0 & 0 & 0 & 0 & -1 & 0 \\
0 & 0 & 0 & 0 & 1 & 0 & 0 \\
0 & 0 & 0 & 1 & 0 & 0 & 0
\end{array}
\right)
\qquad
 Q_2 =\left(
\begin{array}{ccccccc}
0 & 0 & 0 & 0 & 0 & 0 & 0 \\
0 & 0 & 0 & 0 & 0 & 0 & 0 \\
0 & 0 & 0 & 0 & 0 & 0 & 0 \\
0 & 0 & 0 & 0 & 0 & 1 & 0 \\
0 & 0 & 0 & 0 & 0 & 0 & -1 \\
0 & 0 & 0 & -1 & 0 & 0 & 0 \\
0 & 0 & 0 & 0 & 1 & 0 & 0
\end{array}
\right)
$$
$$
Q_3 =\left(
\begin{array}{ccccccc}
0 & 0 & 0 & 0 & 0 & 0 & 0 \\
0 & 0 & 0 & 0 & 0 & 0 & 0 \\
0 & 0 & 0 & 0 & 0 & 0 & 0 \\
0 & 0 & 0 & 0 & -1 & 0 & 0 \\
0 & 0 & 0 & 1 & 0 & 0 & 0 \\
0 & 0 & 0 & 0 & 0 & 0 & -1 \\
0 & 0 & 0 & 0 & 0 & 1 & 0
\end{array}
\right)
\qquad
Q_4 =\left(
\begin{array}{ccccccc}
0 & 0 & 0 & 0 & 0 & 0 & 0 \\
0 & 0 & 0 & 0 & 0 & 0 & 1 \\
0 & 0 & 0 & 0 & 0 & 1 & 0 \\
0 & 0 & 0 & 0 & 0 & 0 & 0 \\
0 & 0 & 0 & 0 & 0 & 0 & 0 \\
0 & 0 & 1 & 0 & 0 & 0 & 0 \\
0 & 1 & 0 & 0 & 0 & 0 & 0
\end{array}
\right)
$$
$$
Q_5 =\left(
\begin{array}{ccccccc}
0 & 0 & 0 & 0 & 0 & 0 & 0 \\
0 & 0 & 0 & 0 & 0 & -1 & 0 \\
0 & 0 & 0 & 0 & 0 & 0 & 1 \\
0 & 0 & 0 & 0 & 0 & 0 & 0 \\
0 & 0 & 0 & 0 & 0 & 0 & 0 \\
0 & -1 & 0 & 0 & 0 & 0 & 0 \\
0 & 0 & 1 & 0 & 0 & 0 & 0
\end{array}
\right)
\qquad
Q_6 =\left(
\begin{array}{ccccccc}
0 & 0 & 0 & 0 & 0 & 0 & 0 \\
0 & 0 & 0 & 0 & 1 & 0 & 0 \\
0 & 0 & 0 & -1 & 0 & 0 & 0 \\
0 & 0 & -1 & 0 & 0 & 0 & 0 \\
0 & 1 & 0 & 0 & 0 & 0 & 0 \\
0 & 0 & 0 & 0 & 0 & 0 & 0 \\
0 & 0 & 0 & 0 & 0 & 0 & 0
\end{array}
\right)
$$
$$
Q_7 =\left(
\begin{array}{ccccccc}
0 & 0 & 0 & 0 & 0 & 0 & 0 \\
0 & 0 & 0 & -1 & 0 & 0 & 0 \\
0 & 0 & 0 & 0 & -1 & 0 & 0 \\
0 & -1 & 0 & 0 & 0 & 0 & 0 \\
0 & 0 & -1 & 0 & 0 & 0 & 0 \\
0 & 0 & 0 & 0 & 0 & 0 & 0 \\
0 & 0 & 0 & 0 & 0 & 0 & 0
\end{array}
\right)
\qquad
Q_8 =\frac 1{\sqrt 3} \left(
\begin{array}{ccccccc}
0 & 0 & 0 & 0 & 0 & 0 & 0 \\
0 & 0 & -2 & 0 & 0 & 0 & 0 \\
0 & 2 & 0 & 0 & 0 & 0 & 0 \\
0 & 0 & 0 & 0 & 1 & 0 & 0 \\
0 & 0 & 0 & -1 & 0 & 0 & 0 \\
0 & 0 & 0 & 0 & 0 & 0 & -1 \\
0 & 0 & 0 & 0 & 0 & 1 & 0
\end{array}
\right)
$$
$$
Q_9 =\frac 1{\sqrt 3} \left(
\begin{array}{ccccccc}
0 & -2 & 0 & 0 & 0 & 0 & 0 \\
2 & 0 & 0 & 0 & 0 & 0 & 0 \\
0 & 0 & 0 & 0 & 0 & 0 & 0 \\
0 & 0 & 0 & 0 & 0 & 0 & 1 \\
0 & 0 & 0 & 0 & 0 & -1 & 0 \\
0 & 0 & 0 & 0 & 1 & 0 & 0 \\
0 & 0 & 0 & -1 & 0 & 0 & 0
\end{array}
\right)
\qquad
Q_{10} =\frac 1{\sqrt 3} \left(
\begin{array}{ccccccc}
0 & 0 & -2 & 0 & 0 & 0 & 0 \\
0 & 0 & 0 & 0 & 0 & 0 & 0 \\
2 & 0 & 0 & 0 & 0 & 0 & 0 \\
0 & 0 & 0 & 0 & 0 & -1 & 0 \\
0 & 0 & 0 & 0 & 0 & 0 & -1 \\
0 & 0 & 0 & 1 & 0 & 0 & 0 \\
0 & 0 & 0 & 0 & 1 & 0 & 0
\end{array}
\right)
$$
$$
 Q_{11} = \frac 1{\sqrt 3} \left(
\begin{array}{ccccccc}
0 & 0 & 0 & -2 & 0 & 0 & 0 \\
0 & 0 & 0 & 0 & 0 & 0 & -1 \\
0 & 0 & 0 & 0 & 0 & 1 & 0 \\
-2 & 0 & 0 & 0 & 0 & 0 & 0 \\
0 & 0 & 0 & 0 & 0 & 0 & 0 \\
0 & 0 & 1 & 0 & 0 & 0 & 0 \\
0 & -1 & 0 & 0 & 0 & 0 & 0
\end{array}
\right)
\qquad
Q_{12} = \frac 1{\sqrt 3} \left(
\begin{array}{ccccccc}
0 & 0 & 0 & 0 & -2 & 0 & 0 \\
0 & 0 & 0 & 0 & 0 & 1 & 0 \\
0 & 0 & 0 & 0 & 0 & 0 & 1 \\
0 & 0 & 0 & 0 & 0 & 0 & 0 \\
-2 & 0 & 0 & 0 & 0 & 0 & 0 \\
0 & 1 & 0 & 0 & 0 & 0 & 0 \\
0 & 0 & 1 & 0 & 0 & 0 & 0
\end{array}
\right)
$$
$$
Q_{13} = \frac 1{\sqrt 3} \left(
\begin{array}{ccccccc}
0 & 0 & 0 & 0 & 0 & -2 & 0 \\
0 & 0 & 0 & 0 & -1 & 0 & 0 \\
0 & 0 & 0 & -1 & 0 & 0 & 0 \\
0 & 0 & -1 & 0 & 0 & 0 & 0 \\
0 & -1 & 0 & 0 & 0 & 0 & 0 \\
-2 & 0 & 0 & 0 & 0 & 0 & 0 \\
0 & 0 & 0 & 0 & 0 & 0 & 0
\end{array}
\right)
\qquad
Q_{14} = \frac 1{\sqrt 3} \left(
\begin{array}{ccccccc}
0 & 0 & 0 & 0 & 0 & 0 & -2 \\
0 & 0 & 0 & 1 & 0 & 0 & 0 \\
0 & 0 & 0 & 0 & -1 & 0 & 0 \\
0 & 1 & 0 & 0 & 0 & 0 & 0 \\
0 & 0 & -1 & 0 & 0 & 0 & 0 \\
0 & 0 & 0 & 0 & 0 & 0 & 0 \\
-2 & 0 & 0 & 0 & 0 & 0 & 0
\end{array}
\right)
$$
\normalsize
The matrices $\{ Q_1 \ , Q_2 \ , Q_3  \ , Q_8 \ , Q_9 \ , Q_{10} \}$ generate the Lie algebra of $SO(4)$,
and the elements $Q_5$ and $Q_{11}$ commute, and generate a non compact Cartan subalgebra contained in $\mathfrak {p}$.

\subsection{Euler construction of \boldmath{$G_{2(2)}/SO(4)$}}
As we have seen, a first way to realize the non compact form is by analytic 
continuation. In this case it simply means that we have to
substitute the matrices $C_I$ with $Q_I$ in (\ref{so4param}):
\begin{eqnarray}
g[x_1,\ldots,x_{14}]=H(x_1,\ldots,x_6) e^{\sqrt 3 x_7 Q_{11} +x_8 Q_5} H(x_9,\ldots,x_{14}), \\
H(x_1,\ldots,x_6)=e^{x_1 Q_3} e^{x_2 Q_2} e^{x_3 Q_3} e^{\sqrt 3 x_4 Q_8} e^{\sqrt 3 x_5 Q_9} e^{\sqrt 3 x_6 Q_8}.
\end{eqnarray}
We can then proceed with the computation of the invariant measure exactly
as for the compact space. We are going to skip all details here and give 
only the final result:
\eqn
d\mu_{G_{2(2)}}=27 \sqrt 3 f(2 x_7 \ , 2x_8 ) \sin (2x_2 )
\sin (2x_5) \sin (2x_{10}) \sin (2x_{13} )
\prod_{i=1}^{14} dx_i \ , \label{invmeasure1}
\feqn
where
\begin{equation}
f(\alpha, \beta )=\sinh (\frac {\beta-\alpha}2 ) 
\sinh (\frac {\beta+\alpha}2 )
\sinh (\frac {\beta-3\alpha}2 ) \sinh (\frac {\beta+3\alpha}2 ) 
\sinh (\alpha) \sinh (\beta),
\label{ffunct1}
\end{equation}
{from} which we can determine the range for the parameters:

\eqn
&& 0 \leq a_1 \leq \pi \ , \qquad  0 \leq a_2 \leq
\frac \pi2 \ ,
\qquad 0 \leq a_3 \leq \frac \pi2 \ , \cr
&& 0 \leq a_4 \leq 2 \pi \ , \qquad  0 \leq a_5 \leq \frac \pi4 \ ,
\qquad 0 \leq a_6 \leq \pi \ , \cr
&& 0 \leq a_9 \leq 2 \pi \ , \qquad  0 \leq a_{10} \leq \frac \pi2 \ ,
\qquad 0 \leq a_{11} \leq \pi \ , \cr
&& 0 \leq a_{12} \leq  \pi \ , \qquad  0 \leq a_{13} \leq \frac \pi2 \ ,
\qquad 0 \leq a_{14} \leq \pi \ , \cr
&& 0 \leq a_7 \leq \infty \ , \qquad  3a_7 \leq a_8 \leq \infty \ .
\feqn
Next, we can also compute the metric (\ref{quoz}) on the quotient $G_{2(2)}/SO(4)$. The details are very similar to the ones in
\cite{CCdVOS}.
Introducing the $1$--forms
\eqn
&& I_1(x,y,z):=\sin(2y) \cos (2z) dx -\sin (2z) dy , \cr
&& I_2(x,y,z):=\sin(2y) \sin (2z) dx +\cos (2z) dy , \cr
&& I_3(x,y,z):=dz+\cos(2y)dx,
\feqn
we get
{\small
\eqn
&& ds^2_{G_{2(2)}/SO(4)} =
da_8^2 +da_7^2 +\left[ \sinh^2 a_8 \cosh^2 a_7 +\cosh^2 a_8 \sinh^2 a_7 \right]
\left( da_5^2 +\sin^2 (2a_5) da_4^2  \right. \cr
&& \qquad \ \left. +3da_2^2 +3\sin^2 2a_2 da_1^2 \right)\cr
&& \qquad \ +\frac 12 \cosh (2a_8) \cosh (2a_7) \sinh^2 (2a_7)
\left\{ \left[ I_1 (a_4 , a_5 , a_6 ) +3 I_2 (a_1 , a_2 ,a_3 ) \right]^2
\right. \cr
&& \qquad \ \left. +\left[ I_2 (a_4 , a_5 , a_6 ) -3 I_1 (a_1 , a_2 ,
a_3 ) \right]^2 \right\} \cr
&& \qquad \ +\frac 34 \sinh^2 (2a_7)
\left[ I_3 (a_4 , a_5 , a_6 ) -I_3 (a_1 , a_2 ,a_3 ) \right]^2 \cr
&& \qquad \ +\frac 14 \sinh^2 (2a_8)
\left[ I_3 (a_4 , a_5 , a_6 ) +3I_3 (a_1 , a_2 ,a_3 ) \right]^2 \ .
\label{hmetric}
\feqn
}
Such a computation is already quite complicated for the $G_2$ group, and
for higher dimensional groups it quickly becomes prohibitive.

\subsection{Iwasawa construction of \boldmath{$G_{2(2)}/SO(4)$}}

The Iwasawa parametrization is the most suitable for the computation of the metric on $G_{2(2)}/SO(4)$.
We know that the Cartan subalgebra of $\mathfrak{p}$ is generated by $H_1:=C_{11}$ and $H_2:=C_{5}$.
The roots of $G_2$ can thus been computed by diagonalizing the adjoint action of $H_i$.  We obtain
\begin{eqnarray*}
&& r_1=(\frac 2{\sqrt 3}, 0); \qquad\ r_2=(\sqrt 3, 1); \qquad\ r_3=(\frac 1{\sqrt 3},1);\\
&& r_4=(0,2); \qquad\ r_5=(-\frac 1{\sqrt 3},1); \qquad\ r_6=(-\sqrt 3, 1),
\end{eqnarray*}
where we write only a choice of positive roots.
The corresponding eigenmatrices (up to some normalization constants) are
\begin{eqnarray}
&& R_1=\sqrt 3 C_3 -C_8 +2C_{12};\\
&& R_2=\frac 1{\sqrt 3} (C_1-C_2+C_6-C_7)-C_9+C_{10}+C_{13}+C_{14};\\
&& R_3={\sqrt 3} (C_1+C_2+C_6+C_7)+C_9+C_{10}-C_{13}+C_{14};\\
&& R_4=C_3-2C_4+\sqrt 3 C_8;\\
&& R_5=-{\sqrt 3} (C_1-C_2+C_6-C_7)-C_9+C_{10}+C_{13}+C_{14};\\
&& R_6=-\frac 1{\sqrt 3} (C_1+C_2+C_6+C_7)+C_9+C_{10}-C_{13}+C_{14}.
\end{eqnarray}
If we choose to parameterize the nilpotent subgroup $N$ as
$N(x_1,\ldots,x_6)=\prod_{i=1}^6 e^{x_i R_i}$, we get
\begin{eqnarray}
&& n^1=dx_1;\\
&& n^2=dx_2-4\sqrt 3 x_1 dx_3+16 x_1^2 dx_5
-\frac{64}{3 \sqrt 3} x_1^3 dx_6;\\
&& n^3=dx_3-\frac 8{\sqrt 3} x_1 dx_5+\frac{16}{3} x_1^2 dx_6;\\
&& n^4=dx_4+8x_3 dx_5 -\frac 83 x_2 dx_6;\\
&& n^5=dx_5-\frac{4}{\sqrt 3}x_1 dx_6;\\
&& n^6=dx_6.
\end{eqnarray}
As we see, these are polynomials.
Then
\eqn
d\sigma^2= dy_1^2+dy_2^2+\sum_{i=1}^6 e_i \otimes e_i
\feqn
with
{\small
\begin{eqnarray*}
&& e^1=-2e^{-2y_2} \left(dx_4+8x_3 dx_5 -\frac 83 x_2 dx_6\right),\\
&& e^2=\frac 1{\sqrt 3} \left(e^{-\sqrt 3 y_1- y_2}  (dx_2-4\sqrt 3 x_1 dx_3
+16 x_1^2 dx_5-\frac{64}{3 \sqrt 3} x_1^3 dx_6)
-e^{\sqrt 3 y_1- y_2} dx_6 \right)\cr
&& \qquad\   +\sqrt 3 \left(e^{-\frac 1{\sqrt 3} y_1- y_2}
(dx_3-\frac 8{\sqrt 3} x_1 dx_5+\frac{16}{3} x_1^2 dx_6)
-e^{\frac 1{\sqrt 3} y_1- y_2} (dx_5-\frac{4}{\sqrt 3}x_1 dx_6)\right)\\
&& e^3= -\frac 1{\sqrt 3} \left(e^{-\sqrt 3 y_1-y_2} (dx_2-4\sqrt 3 x_1 dx_3
+16 x_1^2 dx_5-\frac{64}{3 \sqrt 3} x_1^3 dx_6)
+e^{\sqrt 3 y_1- y_2} dx_6\right) \cr
&& \qquad\   +\sqrt 3 \left(e^{-\frac 1{\sqrt 3} y_1- y_2}
(dx_3-\frac 8{\sqrt 3} x_1 dx_5+\frac{16}{3} x_1^2 dx_6)
+e^{\frac 1{\sqrt 3} y_1- y_2} (dx_5-\frac{4}{\sqrt 3}x_1 dx_6)\right)\\
&& e^4=2 e^{-\frac 2{\sqrt 3} y_1} dx_1,\\
&& e^5= e^{-\sqrt 3 y_1- y_2} (dx_2-4\sqrt 3 x_1 dx_3+16 x_1^2 dx_5
-\frac{64}{3 \sqrt 3} x_1^3 dx_6)-e^{\sqrt 3 y_1- y_2} dx_6
\cr
&& \qquad\  +e^{\frac 1{\sqrt 3} y_1- y_2} (dx_5-\frac{4}{\sqrt 3}x_1 dx_6)
-e^{-\frac 1{\sqrt 3}y_1- y_2}
(dx_3-\frac 8{\sqrt 3} x_1 dx_5+\frac{16}{3} x_1^2 dx_6),\\
&& e^6= e^{-\sqrt 3 y_1- y_2} (dx_2-4\sqrt 3 x_1 dx_3+16 x_1^2 dx_5
-\frac{64}{3 \sqrt 3} x_1^3 dx_6+e^{\sqrt 3 y_1- y_2} dx_6\cr
&& \qquad\  +e^{\frac 1{\sqrt 3}y_1- y_2} (dx_5-\frac{4}{\sqrt 3}x_1 dx_6)
)+e^{-\frac 1{\sqrt 3} y_1- y_2}
(dx_3-\frac 8{\sqrt 3} x_1 dx_5+\frac{16}{3} x_1^2 dx_6),\\
&& e^7=dy_1,\\
&& e^8=dy_2.
\end{eqnarray*}
}
The polynomial dependence on the variables makes the computation feasible
by a computer even for higher dimensional groups.


\section{CONCLUSIONS}

We have given a detailed explanation of the methods for
studying the geometry of exceptional Lie groups that we have
first introduced in \cite{CCdVOS} for $G_2$. Indeed, here
we have seen the elementary reasonings which constitute the
basis of our ideas and provide a powerful tool for computing
global parameterizations of Lie groups. Recall that a
parametrization differs from a coordinatization in that it
does not provide a diffeomorphism between the manifold and the
space of parameters. However, a parametrization locally yields
a coordinatization and it is global when it covers
the whole group. This means that, if $\mathcal{R}$ is the space
of parameters and $G$ the group, then the parametrization
$$
p:\mathcal{R} \longrightarrow G
$$
is surjective. In general, however, it cannot be injective.
Indeed, in general group manifolds have a non vanishing
curvature tensor and cannot be globally covered by a single
chart. Nevertheless, a parametrization can be considered good
when it is ``minimal'', in the sense that $\mathcal{R}$ is
the closure of an open local chart.
This means that the bijectivity of $p$ is lost only on a
subset of vanishing measure, which is the boundary
$\partial \mathcal{R}$ of $\mathcal{R}$. In this case we call
the set $\mathcal{R}$ the range of the parameters.\\
In general, for a finite dimensional simple Lie group the true
difficulty lies not so much in constructing a global
parametrization, but rather in determining the range. Here is
where the idea of the generalized Euler
angles comes into play, as it is particularly suitable for
computing the full range of the parameters, since it allows us to
express them in terms of Cartesian products.\\
In particular, we have seen that there are essentially two
methods to determine the range. The first is geometric and
is based on the detailed knowledge of the geometry of the
quotient space of the group and its maximal subgroup. We have
described it for the example of the $SU(3)$-Euler
parametrization of $G_2$, but it can be adopted for the Euler
parametrization of any of the compact classical simple Lie
groups, as for example $SU(N)$ \cite{BCC}. The second method
is topological and can be used when the geometrical information
on the quotient space is lacking or the geometrical method
is not sufficient to fix the range, as for example
for the $SO(4)$-Euler parametrization of $G_2$ \cite{CCdVOS}
or for the parameterizations of $F_4$ \cite{BCCS1} and $E_6$
\cite{BCCS2}.\\
Finally, we have also considered the construction of
non compact Lie groups. In that case the Iwasawa decomposition is
simpler than the Euler one. In particular, we have shown how it 
can be applied to the non compact Lie group $G_{2(2)}$.
The material exposed in the last section is all new.

\section*{Acknowledgments}

B.L.C. would like to thank S. Ferrara and A. Marrani for enlightening
discussions. The work of B.L.C. has been supported
in part by the Director, Office of Science, Office of High Energy and
Nuclear Physics, Division of High Energy Physics of the U.S. Department
of Energy under Contract No. DE-AC02-05CH11231, and in part by NSF grant
10996-13607-44 PHHXM.


\end{document}